\documentclass[pre,twocolumn,reprint,showpacs,showkeys,groupedaddress,preprintnumbers]{revtex4-1}

\usepackage{verbatim}
\usepackage{graphicx}
\usepackage{amsmath}
\usepackage{float}
\usepackage{amsmath,amssymb,amsfonts}
\usepackage{array}
\usepackage{color}
\usepackage{bbold,ulem}
\usepackage{sidecap}  
\usepackage[english]{babel}
\renewcommand{\emph}{\textit}

\begin{document}

\title{Protein Pattern Formation}
\author{Erwin Frey$^{1}$}
\email{frey@lmu.de}
\author{Jacob Halatek$^{1}$}
\author{Simon Kretschmer$^{2}$}
\author{Petra Schwille$^{2}$}

\affiliation{
$^{1}$Arnold-Sommerfeld-Center for Theoretical Physics and Center for NanoScience, Department of Physics, Ludwig-Maximilians-Universit\"at M\"unchen, Theresienstra{\ss}e 37, D-80333 M\"unchen, Germany \\$^{2}$Department of Cellular and Molecular Biophysics, Max-Planck-Institute of Biochemistry, Am Klopferspitz 18, D-82152 Martinsried, Germany}


\vfill

\begin{abstract}

Protein pattern formation is essential for the spatial organization of many intracellular processes like cell division, flagellum positioning, and chemotaxis. A prominent example of intracellular patterns are the oscillatory pole-to-pole oscillations of Min proteins in \textit{E. coli} whose biological function is to ensure precise cell division. Cell polarization, a prerequisite for processes such as stem cell differentiation and cell polarity in yeast, is also mediated by a diffusion-reaction process. More generally, these functional modules of cells serve as model systems for self-organization, one of the core principles of life. Under which conditions spatio-temporal patterns emerge, and how these patterns are regulated by biochemical and geometrical factors are major aspects of current research. Here we review recent theoretical and experimental advances in the field of intracellular pattern formation, focusing on  general design principles and fundamental physical mechanisms.	
\end{abstract}


\maketitle

\section{Introduction}
\label{sec:introduction}

From cellular structures to organisms and populations, biological systems are governed by principles of self-organisation. 
The intricate cycles of autocatalytic reactions that constitute cell metabolism, the highly orchestrated processes of nucleic acid transcription and translation, the replication and segregation of chromosomes, the cytoskeletal assemblies and rearrangements that mechanically drive important cellular processes like cell division and cell motility, the morphogenesis of complex tissue from a single fertilised egg -
all of these processes rely on the generation of structures and gradients based on molecular self-organisation. 
Frequently, the assembly and maintenance of these structures is accompanied by spatial and temporal protein patterning. 

What are the principles underlying self-organising processes that result in protein patterns? 
Though the term `self-organisation' is frequently employed, as it is here, in the context of complex systems, it needs to be emphasised that there is no generally accepted theory of self-organisation that explains how internal molecular processes are able to coordinate the interactions between a system's components such that order and structure emerge. 
The field which has arguably contributed most to a deeper understanding of emergent phenomena is `nonlinear dynamics', especially with concepts such as `catastrophes' \cite{Thom:1983}, `Turing instabilities' \cite{Turing:1952}, and `nonlinear attractors' \cite{Guckenheimer:2013}.
However, although pattern formation and its underlying concepts have found their way into textbooks \cite{Cross_Greenside:Book}, we are far from answering the above question in a comprehensive and convincing way. 
This chapter will highlight some of the recent progress in the field, but also address some of the fascinating questions that remain open. 

In contrast to the conventional representation of pattern--forbabming systems in classical texts, our exposition will be closely tied to the analysis of quantitive models for specific biological systems. 
At first, this might appear to involve a loss of generality.
However, as we will see, only by studying the actual physical processes that give rise to what we call self-organisation will we be able to uncover its key features in the first place. 
These key aspects can then be generalised again by identifying the according processes in other systems. 
Here, we will mainly, but not exclusively, focus on a model for Min protein dynamics, a system of self-organising proteins that is essential for cell division in the bacterium \textit{Escherichia coli}. 
The Min system offers an ideal combination of a broad and rich phenomenology with accessibility to theoretical and experimental analyses on a quantitative level.
As we will see, a major finding from the study of the Min system is the role of mass-conserved interactions and of system geometry in the understanding of self-organised pattern formation.

\section{Intracellular protein patterns}
\label{sec:intracellular_protein_patterns}

The formation of protein patterns and the localisation of protein clusters is a fundamental prerequisite for many important processes in bacterial cells. 
Examples include Min oscillations that guide the positioning of the Z-ring to midcell in \textit{Escherichia coli}, the localisation of chemotactic signalling arrays and the positioning of flagella, as well as chromosome and plasmid segregation. 
In all these examples, experimental evidence supports mechanisms based on reaction-diffusion dynamics. 
Moreover, the central elements of the biochemical reaction circuits driving these processes are P-loop NTPases. 
These proteins are able to switch from an NTP-bound `active' form that preferentially binds to an intracellular interface (membrane or nucleoid) to an inactive, freely diffusing, NDP-bound form in the cytosol.

Interestingly, these types of pattern--forming--mechanisms are not restricted to prokaryotic cells, but are found in eukaryotic cells as well. 
An important example is cell polarisation, an essential developmental process that defines symmetry axes or selects  directions of growth. 
Signalling molecules accumulate in a restricted region of the inner surface of a cell's plasma membrane where they initiate further downstream processes.  
For example, in the yeast \textit{Saccharomyces cerevisiae}, cell polarisation determines the position of a new growth or bud site. 
The central polarity regulator responsible for this process is Cdc42, a small GTPase of the Rho family \cite{Wedlich-Soldner_etal:2003}.
Similarly, cell polarity plays an important role in proper stem cell division \cite{Florian_Geiger:2010} and in plant growth processes such as pollen tube or root hair development \cite{Molendijk_etal:2001, Gu_etal:2003}.
Another intriguing example of self-organised polarisation occurs in the \textit{Caenorhabditis elegans} zygote through the action of mutually antagonistic, so called partitioning-defective (PAR) proteins \cite{Goehring_etal:2011}. 
Moreover, the crucial role of protein pattern formation in animal cell cytokinesis is highlighted by cortical waves of Rho activity and F-actin polymerization, recently observed in frog and starfish oocytes and embryos \cite{Bement_etal:2015}.

Yet another system where protein patterns play an important role is the transport of motor proteins along cytoskeletal filaments. 
We will not elaborate on this system in this review, but would like to note that pattern formation in these systems is based on similar principles as for the other systems. 
For instance,  microtubules are highly dynamic cytoskeletal filaments, which continually assemble and disassemble through the addition and removal of tubulin heterodimers at their ends~\cite{Desai_Mitchison:2003}. 
It was recently shown that traffic jams of molecular motors on microtubules play a key regulatory mechanism for the length control of microtubules \cite{Varga2006, Varga2009, Reese_etal:2011, Melbinger_etal:2012, Reese2014}.

\subsection{MinCDE oscillations in \textit{E. coli}}
\label{sec:min_oscillations}

Proteins of the Min system in the rod-shaped bacterium \textit{E. coli} show pole-to-pole oscillations \cite{Raskin_deBoer:1999a, Raskin_deBoer:1999b, Hu_Lutkenhaus:1999, Lutkenhaus:2007}. A combination of genetic, biochemical, and cell biological studies has identified the following key features of the underlying interaction network:
(1) The ATPase MinD, in its ATP-bound dimeric form, cooperatively binds to the cytoplasmic membrane \cite{Szeto_etal:2002, Hu_Lutkenhaus:2003, Lackner_etal:2003,Mileykovskaya_etal:2003}, and forms a complex with MinC that inhibits Z-ring formation \cite{Hu_etal:1999}. 
(2) MinD then recruits its ATPase Activating Protein (AAP) MinE to the membrane, triggering MinD's ATPase activity and thereby stimulating detachment of MinD from the membrane in its monomeric form \cite{Hu_Lutkenhaus:2001}. 
(3) Subsequently, MinD undergoes nucleotide exchange in the cytosol and rebinds to the membrane \cite{Hu_etal:2002}.
(4) Notably, MinE's interaction with MinD converts it from a latent to an active form, by exposing a sequestered MinD--interaction region as well as a cryptic membrane targeting sequence \cite{Park_etal:2011, Shih_etal:2011}.
\begin{figure}[b]
\includegraphics[width=\linewidth]{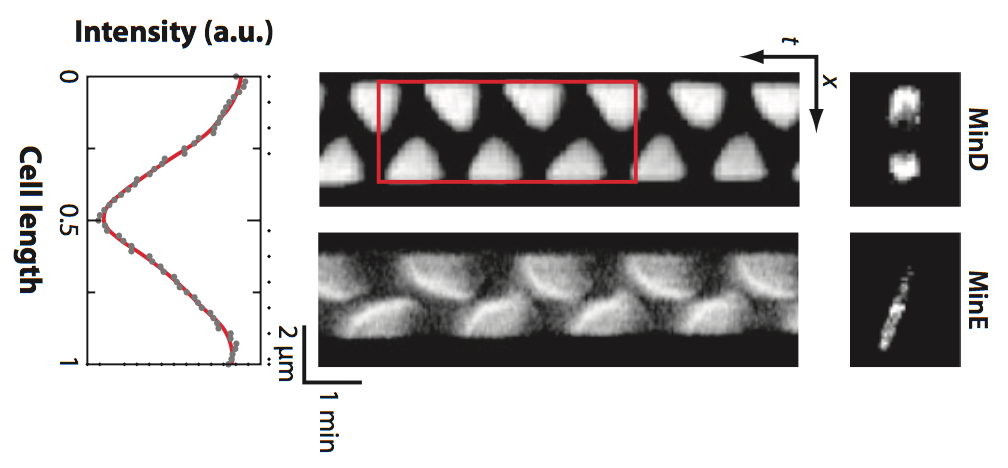}
\caption{\textbf{Oscillatory patterns of Min proteins \textit{in vivo}.} \textit{Left:} Time-averaged MinD fluorescence intensity profile along the red rectangle shown in the kymograph. \textit{Middle:} Kymograph of pole-to-pole oscillations of MinD and MinE in cells of normal length (shorter than $5 \, \mu$m). \textit{Right:} Micrographs of GFP-MinD and MinE-GFP \textit{in vivo}. Adapted from Ref. \protect\cite{Loose_etal:2011_review}. }
\label{fig:exp_min_kymograph}      
\end{figure}

All of these biochemical features give us highly valuable molecular information, but in themselves they do not suffice to explain the emergent phenomenon of Min oscillations. 
There are basically two unknowns. 
First, the detailed dynamic processes underlying, for example, cooperative membrane binding of MinD, as well as the MinE conformational switch are poorly understood on a mechanistic molecular level. 
At present, one can only speculate on them based on structural data. For example, Hill coefficients have been measured for MinD ATP $(\sim 2)$ and ADP $(\sim 1)$ \cite{Mileykovskaya_etal:2003}, indicating that recruitment may be associated with dimerisation. 
Secondly, and perhaps even more importantly, even if all the details of the molecular processes were known, one would still not know which is responsible to what degree for any specific macroscopic property of the dynamic Min pattern. 
Furthermore, how these processes are affected by changing protein expression levels and cell geometry is unclear, \textit{a priori}.
Both of these obstacles represent major challenges for the field, and can be overcome only by a combined experimental and theoretical approach.

The main biological function of Min oscillations is to regulate formation and positioning of the Z-ring \cite{Lutkenhaus:2007}, comprised of curved, overlapping FtsZ filaments, which interact with a range of accessory proteins that together make up the cytokinetic machinery \cite{Lutkenhaus:2012}. 
The pole-to-pole oscillations of the MinD-ATP/MinC complex result in a time-averaged density profile of MinC that is highest at the cell poles and lowest at midcell. 
Since MinC acts as an antagonist of FtsZ assembly, Min oscillations inhibit Z-ring formation at the poles and restrict it to midcell \cite{Hu_etal:1999}.
How self-organisation into the Z-ring occurs remains unknown and is subject to extensive research \cite{Loose_Mitchison:2014, Denk_etal:2016, Ramirez_etal:2016}.

\begin{widetext}

\begin{SCfigure}
    \includegraphics[width=0.7\linewidth]{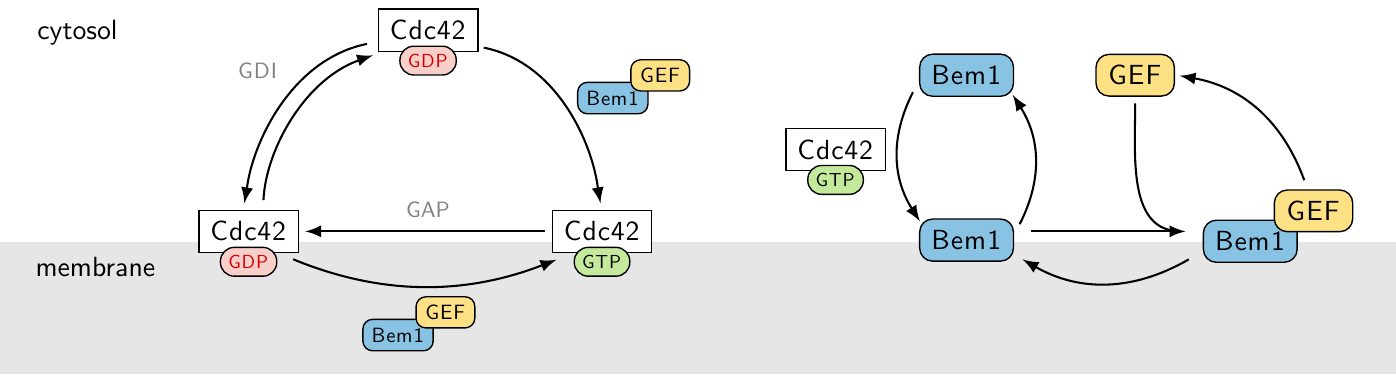}
    \caption{subcaption2}
    \caption{\textbf{Reaction network of the Cdc 42 system in yeast} with a guanine exchange factor (Cdc24) and GAPs controlling the hydrolytic activity of Cdc42. The polarisation relies on activation of Cdc42 through a Bem1-Cdc24-Cla4 complex and on extraction of Cdc42 from membranes by the GDI Rdi1.}
\label{fig:cell_polarity}      
\end{SCfigure}

\end{widetext}

\subsection{Cell polarity in yeast}
\label{sec:cell_polarity}

Polarity establishment in budding yeast relies on crosstalk between feedback loops, one based on the actin cytoskeleton, the other on a reaction-diffusion system \cite{Wedlich-Soldner_etal:2003}. 
Both are regulated by the Rho-type GTPase Cdc42. To fulfil its functions, it must constantly cycle between a GTP-bound (active) and a GDP-bound (inactive) state. 
In budding yeast, activation of Cdc42 is controlled by a single guanine nucleotide exchange factor (GEF), Cdc24, and the hydrolytic activity of Cdc42 is promoted by several GTPase-activating proteins (GAPs). 
In addition, Cdc42 is extracted from membranes by a single Rho-guanine nucleotide dissociation inhibitor (GDI), Rdi1 \cite{Bi_Park:2012}; see Fig. \ref{fig:cell_polarity} for the biochemical network.

Initially two independent feedback loops were identified: one based on the actin cytoskeleton and one based on a reaction-diffusion system that \textit{in vivo} depends on the scaffold protein Bem1 \cite{Bi_Park:2012}. 
A combined experimental and theoretical study has shown that a combination of actin- and GDI-dependent recycling of the GTPase Cdc42 is required to achieve rapid, robust and focused polarisation \cite{Freisinger_etal:2013}. 
However, there are still many open issues on the detailed interplay between these two mechanisms.

The GDI-mediated polarisation in itself is reasonably well understood. 
Theoretical models differ in how they describe the recruitment of the GEF (Cdc24) towards active Cdc42 on the membrane \cite{Goryachev_Pokhilko:2008, Klunder_etal:2013}. 
Experimental data~\cite{Freisinger_etal:2013} support a reaction network where recruitment of Cdc24 is mediated by Bem1  (Fig.~\ref{fig:cell_polarity}): Cytosolic Bem1 is  targeted to the membrane by interaction with active Cdc42 or other Cdc42-GTP-bound proteins such as Cla4 and subsequent binding of Bem1 to the membrane~\cite{Bose_etal:2001, Butty_etal:2002, Kozubowski_etal:2008}. Once bound to the membrane it recruits the Cdc24 from the cytosol to the membrane \cite{Bose_etal:2001, Butty_etal:2002}.
Membrane-bound Cdc24 then enhances both the attachment and activation of cytosolic Cdc42-GDP to the membrane and the nucleotide-exchange of membrane-bound Cdc42-GDP \cite{Freisinger_etal:2013, Klunder_etal:2013}. 
A mathematical model \cite{Klunder_etal:2013} based on this reaction scheme accurately predicts phenotypes associated with changes in Cdc42 activity and recycling, and suggests design principles for polarity establishment through coupling of two feedback loops. 
Recently, there has even been evidence for a third feedback loop~\cite{Bendezu_etal:2015}. 

In a recent \textit{in vivo} study the essential component Bem1 was deleted from the reaction-diffusion feedback loop \cite{Laan_etal:2015}.
 Interestingly, after the mutant was allowed to evolve for about $1,000$  generations, a line was recovered that had regained the ability to polarise, despite the absence of Bem1. 
Moreover, the newly evolved network had actually lost more components, resulting in a simpler reaction-diffusion system.
The structure of this minimal network has yet to be identified \cite{Brauns_etal:unpubl}.

\subsection{Protein pattern formation in animal cell polarisation and cytokinesis}
\label{sec:other_system}

As we have seen for the Min system in \textit{E. coli} and Cdc42 in budding yeast, protein patterns are an elegant way to convey intracellular positional information. 
Thus, it is not surprising that more complex organisms also employ protein pattern formation to control essential processes including cell polarisation, cytokinesis, embryogenesis and development.

An animal's body plan is typically specified during embryogenesis. In this context, the establishment and stable maintenance of cell polarity is a fundamental feature of developmental programs. 
So-called partitioning defective (PAR) proteins are key molecular players that promote symmetry breaking and establish intracellular polarity in diverse animal cells \cite{Goldstein_Macara:2007}. Here, we focus on the PAR network in the nematode worm \textit{C. elegans}, as this system has been particularly well studied.

\begin{widetext}

\begin{SCfigure}
    \includegraphics[width=0.65\linewidth]{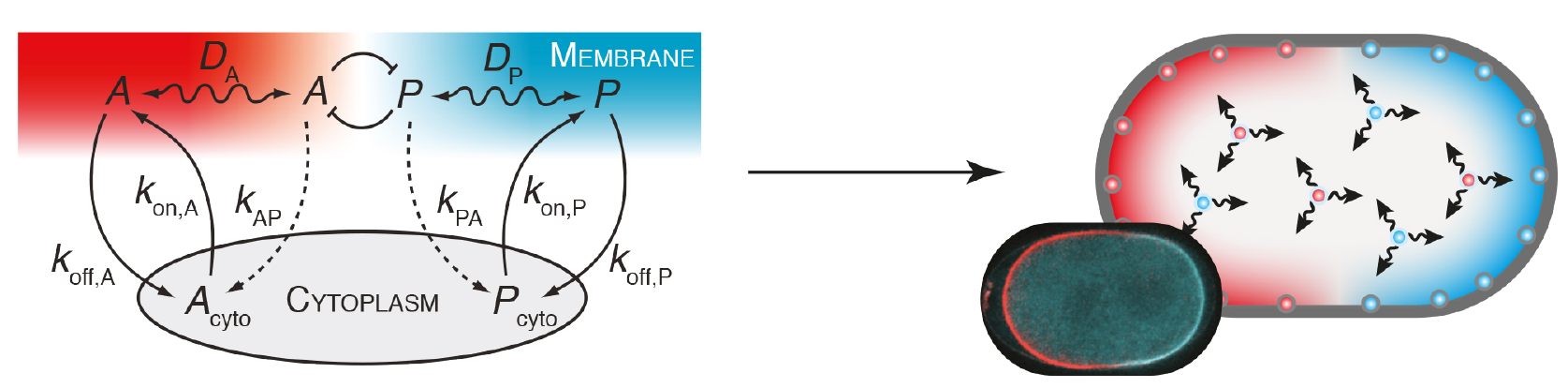}
    \caption{subcaption2}
    \label{fig:2}
\caption{\textbf{Cell polarisation in the \textit{C. elegans} embryo.} A reaction-diffusion network of mutually antagonistic anterior and posterior PAR proteins, switching between ``active'' membrane-bound and ``inactive'' cytosolic states, sustains opposing membrane domains in the  \textit{C. elegans} embryo. Anterior and posterior PAR components are shown in red and blue, respectively. Adapted from reference Ref.~\protect\cite{Goehring_Grill:2013}, copyright 2012 with permission from Elsevier and Ref.~\protect\cite{Goehring_etal:2011} with permission from AAAS.
}
\label{fig:PAR}      
\end{SCfigure}

\end{widetext}

\textit{C. elegans} PAR proteins are required for asymmetric cell division of the  zygote, which they achieve by generating two distinct and complementary membrane domains with the aid of actomyosin flows \cite{Goehring_etal:2011, Munro_etal:2004}.
Several ``design principles'' of the PAR network have been established by a combination of experiments and theory \cite{Goehring:2014}.
A core feature of PAR polarity is the mutual antagonism between anterior and posterior PAR components (Fig. \ref{fig:PAR}), which preferentially accumulate on the anterior and posterior halves of the membrane respectively while being excluded from the opposite half. 

The maintenance of this polarity is highly dynamic and involves mobility of PAR proteins in the cytosol, their cross-inhibition via phosphorylation as well as additional feedback loops \cite{Goehring:2014}. 
Importantly, the mutual antagonism in the PAR network relies on reversible switching of PAR proteins between ``inactive'', rapidly diffusing cytosolic and ``active'', slowly diffusing membrane-bound states \cite{Goehring:2014}, one of the key features of the pattern-forming protein networks discussed in this chapter.

\begin{figure}[b]
\includegraphics[width=\linewidth]{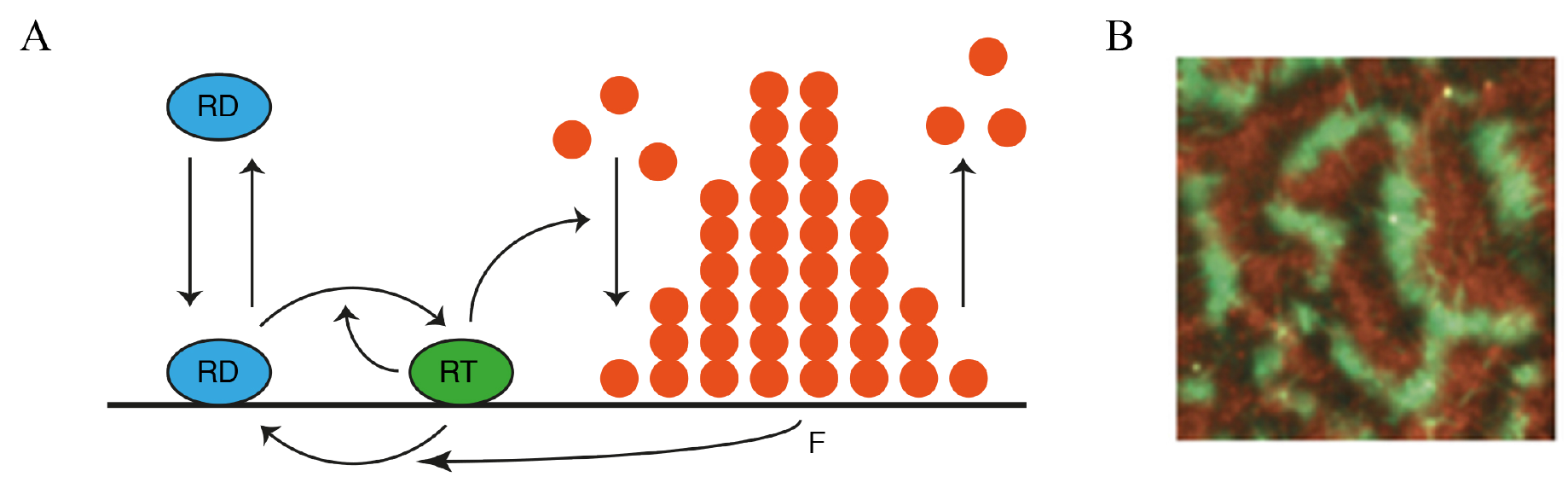}
\caption{\textbf{Cortical waves of Rho activity and F-actin polymerisation involved in animal cell cytokinesis.} 
\textbf{A}, Possible scheme of interactions underlying wave formation. Inactive GDP-bound Rho (RD) binds to the membrane, where it is activated to GTP-bound Rho (RT) via nucleotide exchange in an autocatalytic, GEF-dependent manner. Subsequently, the theoretical model assumes that coupled F-actin polymerisation (F) exerts a negative feedback on Rho activity converting it back into its inactive form \protect\cite{Bement_etal:2015}.
\textbf{B}, Fluorescence image of cortical waves of Rho (malachite) and F-actin (copper) in an Ect2-overexpressing starfish oocyte. Adapted from Ref.~\protect\cite{Bement_etal:2015} by permission from Macmillan Publishers Ltd: Nature Cell Biology \protect\cite{Bement_etal:2015}, copyright 2015.}
\label{fig:starfish}      
\end{figure}

Another intriguing example of protein pattern formation occurs during animal cell cytokinesis. 
This process involves the small GTPase Rho, whose localised activation directs assembly of the cytokinetic machinery, consisting of F-actin and myosin-2, in the equatorial cortex \cite{Green_etal:2012}.
Recently, cortical waves of Rho activity and F-actin polymerisation were discovered in frog and echinoderm oocytes and embryos \cite{Bement_etal:2015}. 
These protein patterns exhibited excitable dynamics and were proposed to emerge through a reaction-diffusion mechanism involving positive feedback during Rho activation and delayed negative feedback exerted by F-actin (Fig. \ref{fig:starfish}). In this view, Rho attaches to the plasma membrane in its inactive GDP-bound form. 
On the membrane, Rho is then converted to its GTP-bound active form in an autocatalytic manner, dependent on the Rho GEF Ect-2. Subsequently, F-actin is assumed to mediate a negative feedback on Rho, converting it back to its inactive form \cite{Bement_etal:2015}. 
Remarkably, this reaction-diffusion network shares many similarities with our previous examples, such as reversible protein attachment to a lipid membrane,  switching between different NTP-bound states and coupling of feedback loops.

\subsection{The switch paradigm}
\label{sec:switch_paradigm}

The molecular mechanisms underlying the spatio-temporal organisation of cellular components in bacteria are frequently linked to P-loop ATPases such as ParA and MinD \cite{Gerdes_etal:2010, Lutkenhaus:2012, Bange_Sinning:2013}. 
ParA and MinD proteins belong to a family of proteins known as the ParA/MinD superfamily of P-loop ATPases \cite{Lutkenhaus:2012}. 
Both are known to form self-organised dynamic patterns at cellular interfaces, ParA on the nucleoid and MinD on the cell membrane. 
The nucleotide state of these ATPases determines their subcellular localisation: While the ATP-bound form dimerises and binds to the appropriate surface, the ADP-bound form  is usually a monomer with a significantly reduced affinity for surface binding that freely diffuses in the cell. 
Importantly, both ParA and MinD have a partner protein (ParB and MinE, respectively) that stimulates their ATPase activity and causes them to detach from their respective surfaces. 
Moreover, there is a delay due to nucleotide exchange between the release of the ADP-bound form from the surface and its subsequent rebinding in the dimeric ATP-bound form.
These interactions enable proteins to cycle between surface-bound and cytosolic states, depending on the phosphorylation state of their bound nucleotide. 
The surface-bound state is typically associated with spatially localised function (e.g.\/ the downstream regulation of other proteins on the surface), whereas the cytosolic state enables spatial redistribution and formation of surface bound patterns of these proteins.
Despite the striking similarities on a molecular level, the biological functions of ParA and MinD differ significantly. 
The Min system directs the placement of the division site at midcell by inhibiting the assembly of FtsZ into a ring-like structure (Z-ring) close to the cell poles. 
In contrast, ParA is involved in chromosome and plasmid segregation. Several other ParA-like proteins have been identified that are also important for the correct localisation of large cellular structures at the cell poles, at midcell or along the cell length \cite{Lutkenhaus:2012}. 
One of these is PomZ in \textit{M. xanthus}. 
PomZ is part of a protein system that -- like the Min system -- is important for Z-ring formation. However, in contrast to the Min system, the Pom system positively regulates the formation of the FtsZ ring at midcell \cite{Treuner-Lange_Sogaard-Andersen:2014} 
Apart from the cell division and the chromosome partitioning machineries, there are various other multiprotein complexes that are positioned by self-organising processes based on P-loop NTPases. 
For example, the GTPase FlhF and the ATPase FlhG constitute a regulatory circuit essential for defining the distribution of flagella in bacterial cells \cite{Bange_Sinning:2013, Schuhmacher_etal:2015}.

\section{Mass-conserving reaction-diffusion systems}
\label{sec:MaRD}

All of the examples of intracellular pattern-forming systems discussed in the previous section share some common features. 
They are reaction-diffusion systems in confined intracellular space, where proteins cycle between the cytosol and the cell membrane \cite{halatek_brauns_frey:2018}.  
On the time scale on which these patterns form, net change in the levels of the proteins involved is negligible and thus the \textit{copy number within each protein species is conserved}. 
The reactions correspond to transitions of each protein species between a finite number of different states (membrane-bound, cytosolic, active, inactive, etc.), and these states play different functional roles in the corresponding biochemical circuit. 
For example, only membrane-bound MinD induces positive and negative feedback by recruiting MinD and MinE from the cytosol to the membrane. 
Hence, the protein dynamics can be understood as a reaction-diffusion system where diffusion takes place in different spatial domains (membrane and cytosol), and where reactions are sequences of state changes induced by protein-nucleotide, protein-protein, and protein-membrane interactions.

\textit{Mass-conserving} dynamics is the generic case for intracellular dynamics. 
Because the production of proteins is a resource-intensive process, any mechanism that utilises production and degradation as pattern forming mechanisms would be highly inefficient and wasteful\footnote{Of course, such a process would also be limited by the duration of protein synthesis.}. 
This excludes activator-inhibitor mechanisms \cite{Segel_Jackson:1972}, since they are based on the interplay between autocatalytic production of a (slow diffusing) activator and its degradation by a (fast diffusing) inhibitor.
Though such a mechanism is frequently invoked as a paradigm in biological pattern formation \cite{Kondo_Miura:2010}, it is actually irreconcilable with the fundamental physical processes on which intracellular pattern formation is based on \cite{halatek_brauns_frey:2018}. 
This in turn implies that the study of biological systems should reveal hitherto unknown mechanisms for pattern formation. Recent research  shows that this is indeed the case \cite{halatek_frey:2018}. In particular, explicit account for mass-conservation yields the total protein densities as system control parameters. As we will see, these are crucial for the theoretical understanding of the experimentally observed phenomena.

\subsection{Cellular geometry: membrane and cytosol}

\begin{figure}[h!]
\includegraphics[width=0.7\linewidth]{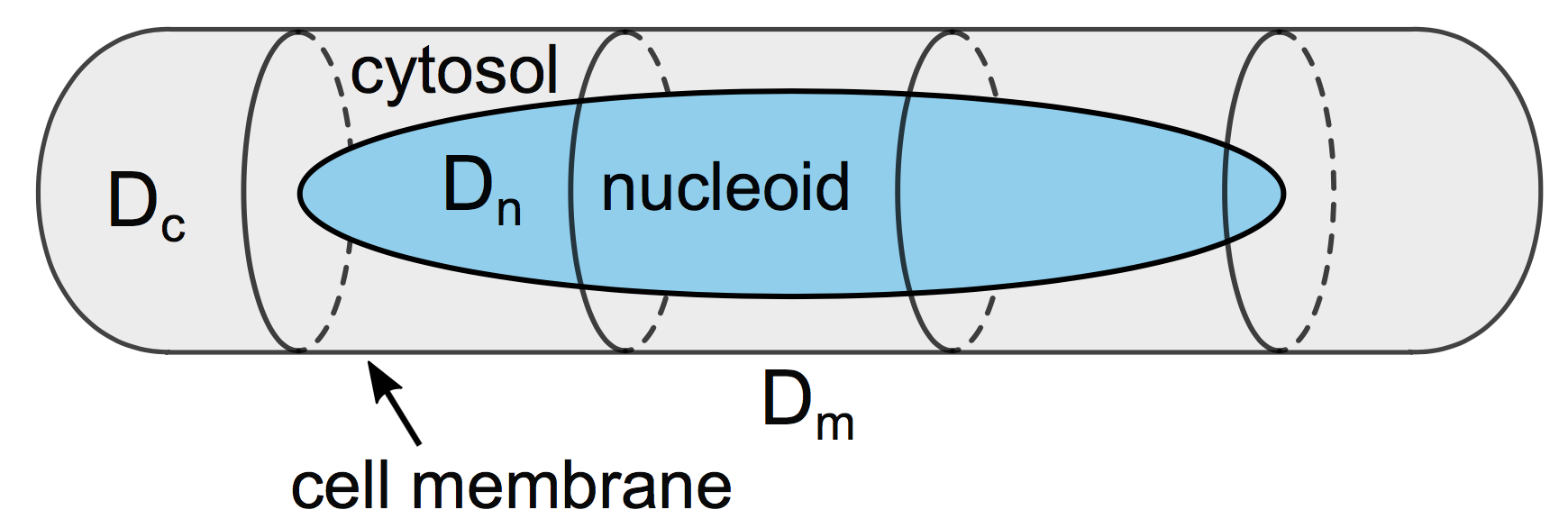}
\caption{\textbf{Schematic representation of the geometry of a rod-shaped bacterial cell.} There are three main compartments: cell membrane, cytosol, and nucleoid. The diffusion constants in these compartments will, in general, be different.
}
\label{fig:cell_geometry}      
\end{figure}

Figure \ref{fig:cell_geometry} illustrates the geometry of a rod-shaped prokaryotic cell. 
It is comprised of three main compartments: the cell membrane, the cytosol, and the nucleoid. 
There are two major facts that are relevant for intracellular pattern formation. 
First, the diffusion constants in the cytosol and on the cell membrane are vastly different. For example, currently accepted values for Min proteins in \textit{E. coli} are of the order of $D_c \,{\approx}\, 10 \, \mu$m$^2$/s, and  $D_m \,{\approx}\,  0.01 \, \mu$m$^2$/s, respectively. 
Second, due to the rod-like shape, the ratio of cytosolic volume to membrane area differs markedly between polar and midcell regions. 
Beyond this local variation of volume to surface ratio, the overall ratio of cytosol volume to membrane area depends on the shape of the cell.

\subsection{Reaction-diffusion equations for the Min system}
\label{sec:MaRD_Min}

\begin{figure}[b]
\includegraphics[width=0.75\linewidth]{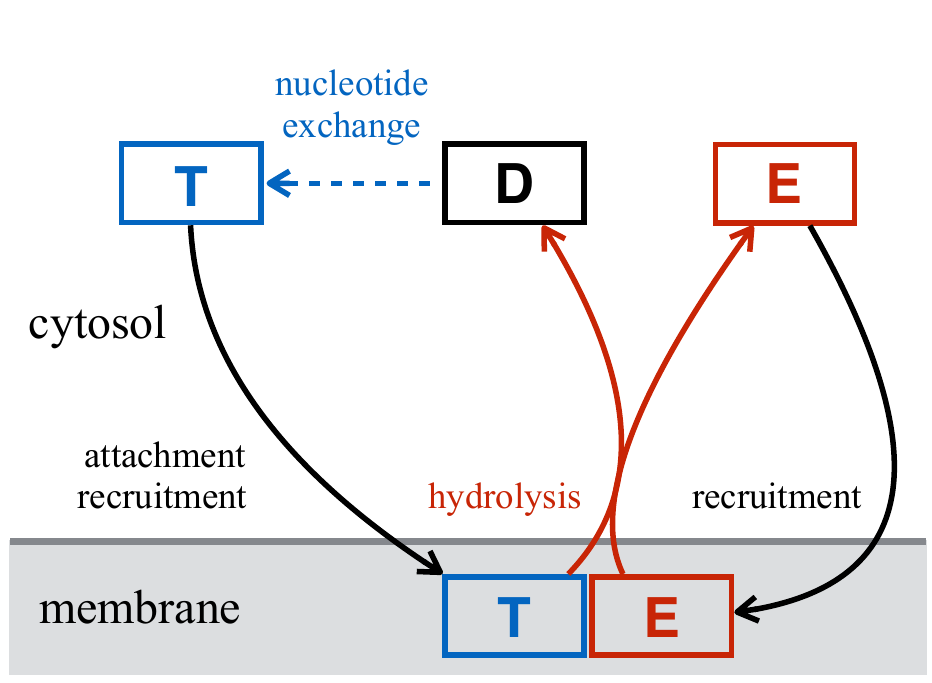}
\caption{\textbf{Skeleton MinCDE network:} Cytosolic MinD-ATP (T) attaches to the membrane, and recruits MinD-ATP and MinE (E) from the cytosol. Recruitment of MinE leads to the formation of MinDE complexes. MinE in the MinDE complexes stimulates ATP hydrolysis by MinD and thereby triggers detachment and dissociation of membrane-bound MinDE complexes into cytosolic MinD-ADP (D) and MinE.
}
\label{fig:min_de_network}      
\end{figure}

The biochemical reactions of the Min system outlined in section \ref{sec:min_oscillations} are summarised in Fig.~\ref{fig:min_de_network}. 
In the following we will refer to this scheme as the \textit{skeleton network}, as it accounts only for those molecular interactions that are (presently) believed to be essential for Min protein phenomenology. 
For a quantitative analysis, this skeleton biochemical network has to be translated into a mathematical model~\cite{Huang_etal:2003, Halatek_Frey:2012}.  

We denote the volume concentrations of MinE, MinD-ADP, and MinD-ATP in the cytosol by $c_{E}^{}$, $c_{DD}^{}$, and $c_{DT}^{}$. 
Since the only reaction that takes place in the cytosol is reactivation of cytosolic MinD-ADP by nucleotide exchange (with rate $\lambda$) to MinD-ATP, the ensuing reaction-diffusion equations read:
\begin{subequations}
\begin{align}
	\partial_{t}c_{DD}^{} 
	&= D_{c}\nabla^{2} c_{DD}^{} -
	   \lambda \, c_{DD}^{}  \, ,
	\label{eq:de1} \\
		\partial_{t}c_{DT}^{} 
	&= D_{c}\nabla^{2}c_{DT}^{} + 
	   \lambda \, c_{DD}^{}  \, ,
	\label{eq:de2}\\
	\partial_{t}c_{E}^{}  
	&= D_{c}\nabla^{2}c_{E}^{}  \, ,
	\label{eq:de3}
\end{align}	
\label{eq:RD_cytosol}
\end{subequations}
The diffusion coefficients are typically distinct for all protein configurations, for simplicity, we only distinguish between cytosolic $(D_c)$ and membrane bound $(D_m)$ states.

Only the active form of MinD, $c_{DT}$, can attach to the membrane, either spontaneously with a rate $k_D$ or facilitated by MinD-ATP already bound to the membrane (recruitment) with a rate $k_{dD}^{} m_{d}^{}$, where $m_{d}^{}$ denotes the areal density of MinD-ATP on the membrane. 
Overall then, the reaction term reads $R_{D}^+ = (k_{D}^{} + k_{dD}^{} \, m_{d}^{}) \, \tilde c_{DT}^{}$, where the tilde on the cytosolic concentration of MinD-ATP indicates that the value must be taken in the immediate vicinity of the membrane. 
Membrane bound MinD-ATP can also recruit cytosolic MinE to the membrane and thereby form MinDE complexes. 
The corresponding reaction term reads $R_{E}^+ = k_{dE}^{} \, m_{d}^{} \, \tilde c_{E}^{}$.  
Finally, MinE in the MinDE complexes stimulates ATP hydrolysis by MinD and hence facilitates detachment and decay of membrane bound MinDE complexes into cytosolic MinD-ADP and MinE, $c_{E}^{}$, with rate $k_{de}^{}$.
This process is described by the reaction term $R_{DE}^- = k_{de}^{} \, m_{de}^{}$ where $m_{de}^{}$ denotes the areal density of MinDE complexes on the membrane.
Taken together, the reaction-diffusion equations on the membrane read
\begin{subequations}
\begin{align}
	\partial_{t}m_{d}^{} 
	&= D_{m}\nabla_{m}^{2}m_{d}^{} + 
	   R_{D}^+ (m_{d}^{}, \tilde c_{DT}^{}) - 
	   R_{E}^+(m_{d}^{}, \tilde c_{E}^{}), 
	\label{eq:de4}\\
	\partial_{t}m_{de}^{} 
	&= D_{m}\nabla_{m}^{2}m_{de}^{} + 
	   R_{E}^+(m_{d}^{}, \tilde c_{E}^{}) -
	   R_{DE}^-(m_{de}^{})  \, ,
	\label{eq:de5}
\end{align}
\label{eq:RD_membrane}
\end{subequations}
where the index $m$ denotes the Laplacian for membrane diffusion.

These two sets of reaction-diffusion equations, Eq.~\ref{eq:RD_cytosol} and Eq.~\ref{eq:RD_membrane}, are complemented by nonlinear reactive boundary conditions at the membrane surface that guarantee local particle number conservation. 
In other words, the chemical reactions involving both membrane-bound and cytosolic proteins equal the diffusive flux onto $(-)$ and off $(+)$ the membrane (the index $\perp$ denoting the outward normal vector at the boundary):
\begin{subequations}
\begin{align}
	\left. D_{c}\nabla_{\perp} c_{DD}^{}\right|_{m} 
	& = + R_{DE}^-(m_{de}^{}) 
	  \, , 
	\label{eq:bc1}\\
	\left. D_{c}\nabla_{\perp} c_{DT}^{}\right|_{m} 
	& =  - R_{D}^+ (m_{d}^{}, \tilde c_{DT}^{}) \, ,
	\label{eq:bc2}\\
	\left. D_{c}\nabla_{\perp} c_{E}^{}\right|_{m} 
	& =  + R_{DE}^-(m_{de}^{})  - 
	     R_{E}^+(m_{d}^{}, \tilde c_{E}^{})
	  \, .
	\label{eq:bc3}
\end{align}
\label{eq:RD_boundary}
\end{subequations}
For example, Eq.~\ref{eq:bc1} states that detachment of MinD-ADP following hydrolysis on the membrane is balanced by gradients of MinD-ADP in the cytosol. 
In general, any exchange of proteins between the membrane and cytosol leads to diffusive fluxes and thereby to protein gradients in the cytosol since the membrane effectively acts as a sink or source of proteins. 
These gradients are essential for understanding the mechanisms underlying intracellular pattern formation, and preclude a naive interpretation of the cytosol as a spatially uniform reservoir. 

For the model to be complete, one needs to know the values of all of the reaction rates. 
However, the estimation and choice of system parameters is a highly nontrivial problem.
Nonlinear systems are generically very sensitive to parameter changes, whereas biological function has to be sufficiently robust against variations in the kinetic rates and diffusion coefficients (e.g.\/ caused by temperature changes).
In addition, only rarely are the system parameters known quantitatively from experiments.
For the Min system only the diffusion coefficients have been measured and estimates for the nucleotide exchange rate $\lambda$ \cite{Meacci_etal:2006} and the Min protein densities exist \cite{Shih_etal:2002}.
However, a theoretical investigation of the skeleton model by means of linear stability analysis and numerical simulations was able to identify parameter regimes where the experimentally observed patterns are formed \cite{Halatek_Frey:2012}. 

\subsection{Basic mechanisms underlying Min oscillations in \textit{E. coli} cells}

From the analysis of the skeleton model~\cite{Halatek_Frey:2012}, quantified by the reaction-diffusion equations in the previous section, one can now learn how Min proteins self-organize to give rise to pole-to-pole oscillations \textit{in vivo}.

The basic theme of the protein dynamics is the cycling of proteins between the membrane and the cytosol. This cycling is driven by the antagonistic roles of MinD and MinE: 
Membrane-bound active MinD facilitates flux of MinD and MinE from the cytosol to the membrane (recruitment). 
This accumulation of proteins at the membrane is counteracted by MinE's stimulation of MinD's ATPase activity, which triggers detachment of both MinD and MinE. 
In concert with redistribution of proteins through cytosolic diffusion, spatio-temporal patterns may emerge on the membrane. 

However, the formation of pole-to-pole oscillations is by no means generic in the context of the above reaction scheme.\footnote{In general, a given reaction-diffusion equation can generate a plethora of spatio-temporal patterns, as is well known from classical equations like the complex Ginzburg-Landau equation \cite{Aranson_Kramer:2002} or the Gray-Scott equation \cite{Gray_Scott:1983, Gray_Scott:1984, Gray_Scott:1985, Pearson:1993, Lee_etal:1993}. Conversely, a given pattern can be produced by a vast variety of mathematical equations. Hence, one must be careful to avoid falling into the trap: ``Cum hoc ergo propter hoc'' (correlation does not imply causation).} 
In general, there are conditions on the values of the reaction rates, as well as on the relative abundances of the proteins which have to be met. 
An exhaustive parameter scan for model equations Eq.~\ref{eq:RD_cytosol}, \ref{eq:RD_membrane}, and \ref{eq:RD_boundary} has shown that, for spatial patterns to emerge in the skeleton model, MinE needs to be recruited faster to the membrane-bound protein layer than MinD, while being lower in total particle number \cite{Halatek_Frey:2012}
\begin{equation}
	k_{dD} < k_{dE} \, , \quad
	N_{E}  < N_{D} \, .
\end{equation}
These conditions give rise to the formation and separation of MinD and MinDE domains, the \textit{polar zone} and \textit{MinE ring}, as the two basic emergent structures of pole-to-pole oscillations. 
\begin{widetext}

\begin{figure}[t]
\centering
\includegraphics[width=\columnwidth]{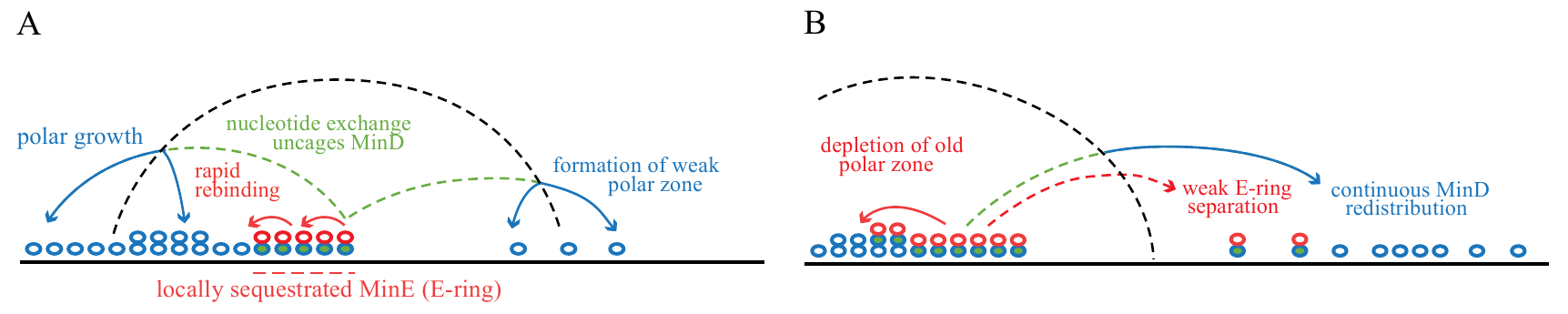}
\caption{\textbf{Key mechanism underlying Min oscillations.}  \textbf{A,} Locally sequestrated MinE constitutes the MinE ring, which moves toward the left pole through local cycling. Detaching MinD rebinds predominantly at the left pole and initiates formation of a weak polar zone at the right end. The delay in reattachment caused by the need for nucleotide exchange is indicated by dashed lines.  \textbf{B,} MinE depletes the old polar zone of MinD, until only MinDE complexes are left, then reassembles at the rim of the new polar zone, formed by redistributed MinD. Adapted from Ref.~\cite{Halatek_Frey:2012} under the CC BY 4.0 license. 
}
\label{fig:min_oscillation_mechanism}      
\end{figure}	

\end{widetext}
As illustrated in Fig.\ref{fig:min_oscillation_mechanism}, this is (heuristically) understood as follows \cite{Halatek_Frey:2012}.
The higher particle number of MinD enables complete sequestration of MinE in membrane-bound MinDE complexes, while leaving a fraction of MinD available to initiate a new polar zone.\footnote{It should be noted that the condition on the particle numbers mainly serves to emphasise the sequestration mechanism. In order for MinD to accumulate in polar zones the action of MinE must be disabled, and specifying that there are fewer MinE particles permits them to be spatially confined. Outside of this zone MinD can accumulate on the membrane. It has been speculated \cite{Halatek_Frey:2012} that other mechanisms, such as transient MinE membrane binding, might provide alternative ways to transiently disable the action of MinE, removing the requirement from the particle numbers. The exact mechanism needs to be investigated in future experiments as well as in the framework of theoretical models.} 
Given a sufficiently high MinD membrane concentration and MinE recruitment rate $k_{dE}$, detaching MinE rebinds immediately, forming the prominent MinE ring.
Continuous MinE cycling locally depletes the membrane of MinD, leading to a slow poleward progression of the MinE ring along the gradient of membrane bound MinD, whereupon a fraction of detaching MinD initiates a weak polar zone in the opposite cell half, see Fig.\ref{fig:min_oscillation_mechanism}A. 
The new polar zone grows due to steady redistribution of MinD, while most MinE remains sequestrated in the old polar zone until the remaining MinD molecules are converted into MinDE complexes, see Fig.\ref{fig:min_oscillation_mechanism}B.
Once this state is reached, the Min proteins rapidly detach, dissociate and diffuse through the cytosol and rapidly reattach at the new polar zone, leaving behind a region of high MinDE/MinD ratio, where immediate reformation of polar zones is inhibited. 
Due to the faster recruitment  of MinE, the MinE ring reassembles at the rim of the new polar zone, which provides the crucial separation of MinD and MinDE maxima, i.e.\/ a polar zone and a MinE ring. 

\begin{widetext}
	
\begin{figure}[t]
\includegraphics[width=0.9\linewidth]{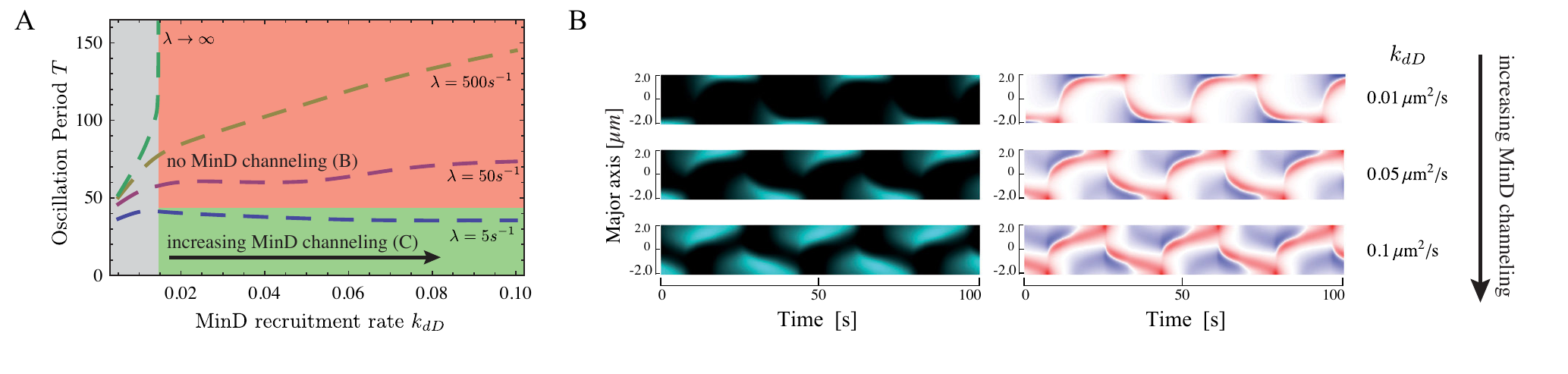}
\caption{\textbf {Canalised MinD transfer and regulation of spatial MinD reattachment by MinD recruitment.} \textbf{A}, Temporal period of Min oscillations as a function of the MinD recruitment rate $k^{}_{dD}$, and nucleotide exchange rate $\lambda$ in cells of $4 \, \mu$m length. With instantaneous nucleotide exchange, oscillations only exist at low MinD recruitment rates (grey). Beyond this threshold the nucleotide exchange and recruitment rates become control parameters for the spatial distribution of MinD reattachment. At high but finite nucleotide exchange rates the oscillation period increases with the MinD recruitment rate, as MinD reassembles in front of the polar zone. At low nucleotide exchange rates the oscillation period decreases with MinD recruitment, as the pole-to-pole particle transfer becomes canalised between the two cell halves. \textbf{B}, Kymographs for $\lambda=5s^{-1}$ showing the total MinD membrane density, $m^{}_d+m^{}_{de}$, and MinD flux $J^{}_D = D^{}_{D}\nabla_{\perp}(c^{}_{DT}+c^{}_{DD})|_{m}$ on (blue) and off (red) the membrane, for a set of increasing MinD recruitment rates $k^{}_{dD}$. MinD reaccumulates at the opposite cell pole while the old pole is still present. Increasing MinD recruitment accelerates the growth of new polar zones towards midcell and synchronises depletion and formation of polar zones at opposite cell ends by canalising the MinD flux from old to new polar zones. Adapted from Ref.~\cite{Halatek_Frey:2012} under the CC BY 4.0 license.
}
\label{fig:min_oscillation_period}      
\end{figure}

\end{widetext}

There is one element of the above argument which needs further consideration: The sequestration of MinE is transient, and hence the system is oscillatory, only if detaching MinD gradually leaks from the old to the new polar zone. But, how is this process established and regulated?
Leakage from the old polar zone is determined by the balance between two opposing factors: the ATPase cycle of MinD, and the propensity of cytosolic MinD to bind to the membrane. MinE stimulates ATPase activity of MinD and thereby initiates detachment of ADP-bound MinD.
The inactive MinD cannot reattach to the membrane until it is reactivated by nucleotide exchange. 
This delay implies that the zone near the membrane is depleted of active MinD, i.e.\/ active MinD has time to diffuse further away from the membrane into the cytosol. 
Taken together, these factors effectively suppress immediate reattachment of MinD and promote its leakage from the polar zone: The slower the nucleotide exchange the more particles leak from polar zones. 
This is counteracted by MinD recruitment: The stronger the recruitment, the ``stickier'' the membrane and hence the fewer particles leak from polar zones. 
Clear evidence for this reasoning comes from the slowing down of the oscillation with increasing nucleotide exchange and MinD recruitment rates, depicted in Fig.\ref{fig:min_oscillation_period}A.

Numerical simulation of the reaction-diffusion equations, Eq.\ref{eq:RD_cytosol}--\ref{eq:RD_boundary}, reveals further functional characteristics of Min oscillations. 
For nucleotide exchange rates $\lambda = 5 \, s^{-1}$, close to the experimentally determined lower bound of $3 \, s^{-1}$, reaccumulation of the polar zone always starts in the opposite cell half, and the recruitment rate $k_{dD}$ of MinD regulates how fast the new polar zone grows towards the old one (Fig. \ref{fig:min_oscillation_period}B).
Notably, at $k_{dD} = 0.1 \, \mu$m$^2$/s in Fig. \ref{fig:min_oscillation_period}B, the redistribution of MinD from old to new polar zone is highly canalised, i.e.\/ the total MinD flux is directed towards the opposite cell half immediately after the polar zones start to shrink (Fig.\ref{fig:min_oscillation_period}B). This implies that growth and depletion of polar zones are synchronised. This is also reflected in the characteristic triangular shape observed in MinD kymographs~\cite{Loose_etal:2011_review}, where new polar zone start growing towards midcell while old polar zones shrink towards the cell pole (Fig. \ref{fig:min_oscillation_period}B). 

Although most of the Min protein patterns (like stripe patterns) observed in filamentous mutant \textit{E. coli} have no biological function, the theory is able to account for their occurrence. This argues strongly that they too arise from the mechanism that optimises the spatial profile of pole-to-pole oscillations for midcell localisation. In other words, the rich phenomenology in mutant cells appears to be a byproduct of the evolutionary optimisation of the wild-type dynamics.

\newpage

\subsection{Cell geometry and pattern formation}

To ensure robustly symmetrical cell division, one would expect Min patterns to scale with cell size and shape, at least within the biologically relevant range. 
Indeed, recent experiments using `cell-sculpting' techniques  \cite{Wu_etal:2015} have shown that longitudinal pole-to-pole oscillations are highly stable in cells with widths below $3\mu$m, and lengths in the range of $3-6 \, \mu$m.
Interestingly, however, outside of this range of cell geometries, Min proteins show diverse oscillation patterns, including longitudinal, diagonal, rotational, striped, and even transverse modes \cite{Raskin_deBoer:1999b, Shih_etal:2005, Wu_etal:2016, Corbin_etal:2002, Touhami_etal:2006, Varma_etal:2008, Maennik_etal:2012, Wu_etal:2015}.
What is the origin of the simultaneous robustness of Min oscillations inside the biologically relevant regime and the bewildering diversity of patterns and multistability outside of it? In what sense are these seemingly contradictory features two faces of the same coin?

To answer these questions one has to address how and to what extent the existence and stability of different patterns is affected by a cell's geometry, and which specific biomolecular processes in the Min reaction circuit control how the system adapts to cell geometry. 
This has recently been achieved by a combination of numerical studies, based on the reaction-diffusion model discussed in section \ref{sec:MaRD}, and experimental studies, in which the geometry of \textit{E. coli} bacteria was systematically varied \cite{Wu_etal:2016}.

There are basically two types of randomness that may affect the process of pattern selection, or transitions between patterns if multiple stable patterns are possible. 
First, the inherent randomness of any chemical reaction may cause stochastic transitions between patterns.
Though such stochastic effects are possible in principle \cite{Fange_Elf:2006}, given the large copy number of Min proteins, they are unlikely to be the major source for transitions between patterns; factors like heterogeneities and asymmetries are expected to be far more important. Second, there are many different factors which cause realistic cellular systems to be asymmetric or heterogeneous. 
For example, the membrane affinity of MinD depends on the lipid composition, which in turn is sensitive to membrane curvature. Hence, small asymmetries of the cell shape translate to variations of MinD membrane attachment.
While these asymmetries and heterogeneities are intrinsic to ensembles of cells, they need to be specifically emulated in numerical simulations. A natural choice are gradients in the MinD attachment rate that are inclined at all possible angles to the long axis of the cell. 
The magnitude of these gradients must be sufficiently large to significantly affect the pattern selection process, but at the same time small enough not to cause any asymmetry in the final stable pattern. 
A relative magnitude of variation in the range of $20 \%$ (well below the natural variability of MinD affinity to different lipids \cite{Mileykovskaya_etal:2003, Renner_Weibel:2012}) fulfills these requirements.

\begin{widetext}

\begin{figure}[bt]
\includegraphics[width=\linewidth]{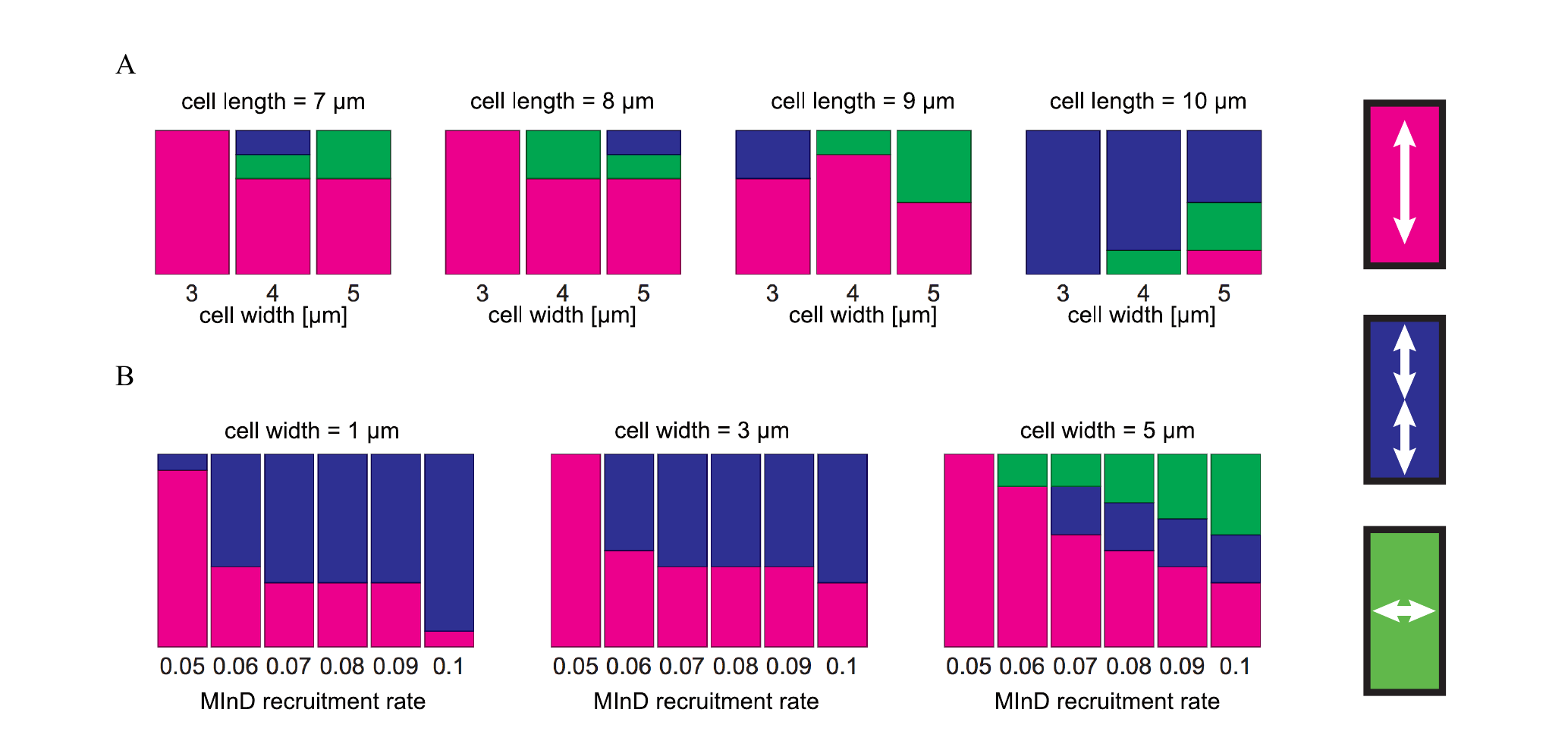}
\caption{\textbf{Basins of attraction predicted from systematic perturbations of patterns with shallow attachment gradients.} 
\textbf{A,} Relative distribution of the final patterns (indicated on the right) observed after sampling all alignment angles of the MinD attachment template from 0 to 90 degrees. The MinD recruitment rate was set to a constant value $k^{}_{dD} = 0.1$. The data shows the increase in the incidence of multistability as the cell size is increased beyond minimal values for cell length and cell width.
\textbf{B,} Fractions of the final patterns in cells of 9- and 10-$\mu$m length after sampling all alignment angles of the MinD attachment template from 0 to 90 degrees. The data shows that increasing the MinD recruitment rate facilitates multistability. Adapted from Ref.~\protect\cite{Wu_etal:2016} under the CC BY 4.0 license. 
}
\label{fig:multistability_theory}      
\end{figure}

\end{widetext}

Figure \ref{fig:multistability_theory} shows histograms of the final stable patterns obtained by sampling over all directions of the gradient, as a function of cell width and length, and of the MinD recruitment rate \cite{Wu_etal:2016}. 
For a recruitment rate fixed to the value that facilitates canalised transfer, $k^{}_{dD} = 0.1$, the following observations are of note. 
(i) As cell length is increased, striped oscillations become more frequent patterns. 
(ii) The fraction of oscillatory striped patterns tends to decrease in favour of transverse patterns as the cell width increases, indicating that cell width, and not cell length, is the main determinant for the onset of transverse modes. 

Both observations are remarkably consistent with  experimental data based on random sampling of live \textit{E. coli} cells after they have reached a defined shape \cite{Wu_etal:2015}. 
Numerical simulations allow us to go beyond the analysis of cell geometry, and investigate the effect of MinD recruitment rate, see Fig.\ref{fig:multistability_theory}B. 
In narrow cells with widths ranging from $1 \, \mu$m to $3 \, \mu$m, one observes that the fraction of stripes increases with the MinD recruitment rate \cite{Halatek_Frey:2012, Wu_etal:2016}. 
In contrast, for cells that reach a width of 5 $\mu$m, stripe patterns are absent below some threshold MinD recruitment rate. 
With increasing MinD recruitment rate, transverse patterns appear first and increase in frequency, while the fraction of striped patterns takes on a constant value.

There are several conclusions one can draw from these observations. 
The most obvious one is that multistability in Min patterns is not determined by either kinetic parameters or cell geometry alone, but originates from the interdependence between these two factors.
In addition, increasing the size of a Turing-unstable system alone does not in itself facilitate the existence of multiple stable patterns \footnote{This is surprising, because Turing instabilities are generically associated with the existence of a \textit{characteristic} (or intrinsic) wave length in the literature. This is evidently not the case here.}. 
This is clearly evident from the observation that the emergence of a pole-to-pole oscillation in a short cell does not generically imply the existence of a stable striped oscillation with a characteristic wavelength in a long filamentous cell \cite{Halatek_Frey:2012}. 
Instead, the emergence of a characteristic length scale (which becomes manifest in striped oscillations) is restricted to a specific regime of kinetic parameters, where growth and depletion of spatially separated polar zones become synchronised such that multiple, spatially separated polar zones can be maintained simultaneously (``canalised transfer'' regime) \cite{Halatek_Frey:2012}.  
A key element among the prerequisites that permit this regime to be reached is that the degree of nonlinearity in the kinetics of the system (MinD cooperativity) must be particularly strong. 
Notably, the same mechanism that enables striped oscillations in filamentous cells also facilitates transverse oscillations in wide cells.

These findings hint at an exciting connection between multistability, the ability of patterns to sense and adapt to changes in system geometry, and the existence of an intrinsic length scale in the underlying reaction-diffusion dynamics. 
Remarkably -- and contrary to the treatments in the classical literature -- the existence of an intrinsic length scale is not generic for a Turing instability per se. 
One example is the aforementioned selection of pole-to-pole patterns in arbitrarily long cells where MinD recruitment is weak. 
In this case, irrespective of the critical wavenumber of the Turing instability, the final pattern is always a single wave traveling from pole to pole. 
The selection of a single polar zone is also characteristic in the context of cell polarity \cite{Klunder_etal:2013, Otsuji_etal:2007}, where it has been ascribed to the finite protein reservoir and a winner-takes-all mechanism. 
It will be an interesting task for further research to elucidate the general requirements for the emergence of an intrinsic length scale in mass-conserved reaction-diffusion systems. 

\subsection{Principles of adaptation to geometry in reaction-diffusion systems}

How does the geometry of a cell affect the formation of spatio-temporal patterns? This question may be rephrased in more mathematical terms as follows: What are the inherent features of a reaction-diffusion system in confined geometry that promote or impede the adaptation of the ensuing patterns to the size and shape of that confining space\footnote{In 1966 Mark Kac published an article entitled ``Can one hear the shape of a drum?''\cite{Kac:1966}. As the dynamics (frequency spectrum) of an elastic membrane whose boundary is clamped is described by the Helmholtz equation $\nabla^2 u + \sigma u =0$ with Dirichlet boundaries, $\nabla u \mid_\perp = 0$, this amounts to asking how strongly the eigenvalues $\sigma$ depend on the shape of the domain boundary. Here we ask a much more intricate question, as the dynamics of pattern forming systems are nonlinear and we would like to know the nonlinear attractor for a given shape and size of a cell.}? In previous sections, we have seen two recurrent themes: nucleotide exchange and positive feedback through recruitment. To elucidate the role of these two factors we will in this section shortly review recent results \cite{Thalmeier_etal:2016} for a minimal pattern-forming system comprised of a single NTPase only.
\begin{figure}[b]
\centering
\includegraphics[width=0.65\linewidth]{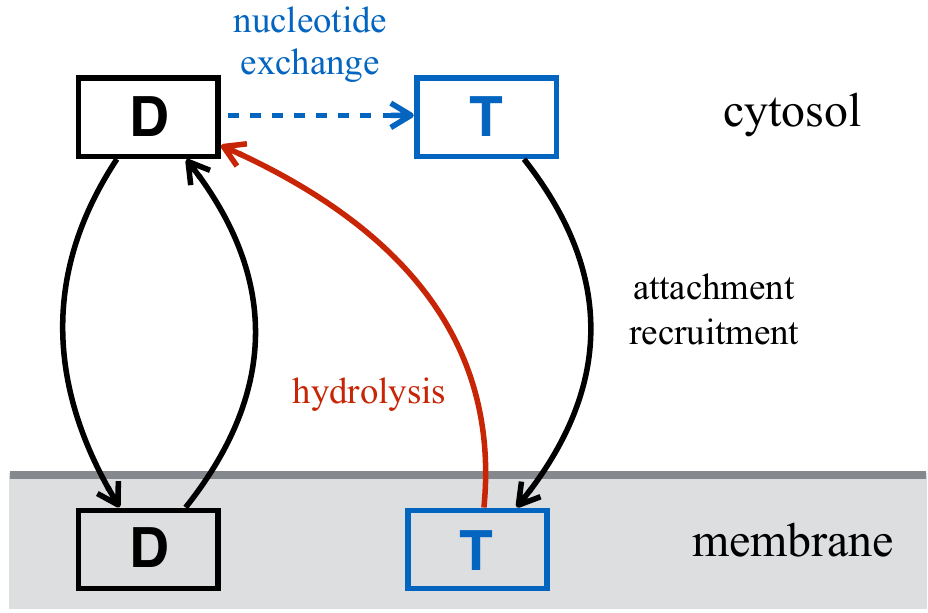}
\caption{
The NTPase can bind to the membrane in both of its states with \textit{attachment rate} $k_+$, or cooperatively with corresponding \textit{recruitment rates} $k_{mD}$ for $D$ and $k_{mT}$ for $T$. NTP \textit{hydrolysis} by $T$ triggers \textit{detachment} with rate $k_-$, converting membrane-bound $T$ into cytosolic $D$. Membrane-bound $D$ is also spontaneously released to the cytosol with \textit{detachment rate} $k_-$. Cytosolic $D$ undergoes \textit{nucleotide exchange} with a rate $\lambda$.
}
\label{fig:one_ntpase_model}      
\end{figure}

As illustrated in Fig.\ref{fig:one_ntpase_model}, the NTPase cycles between an NDP-bound inactive ($D$) and an NTP-bound active state ($T$). 
Both protein species are able to bind to the membrane spontaneously; for simplicity we take the rates to be identical and given by $k_+$.
In addition, to direct membrane attachment, each protein species may also bind cooperatively to the membrane with corresponding recruitment rates $k_{mD}$ for the inactive and $k_{mT}$ for the active protein species. Detachment of the membrane-bound species is asymmetric: While the inactive species is simply released to the cytosol with \textit{detachment rate} $k_-$, detachment of the active species is triggered by NTP hydrolysis which is thereby converted into cytosolic inactive $D$; again, for simplicity, we assume the corresponding detachment rates to be equal and given by $k_-$. Reactivation of cytosolic inactive $D$ through nucleotide exchange occurs at rate rate $\lambda$. Both protein forms are allowed to freely diffuse in the cytosol and the membrane with diffusion constants $D_c$ and $D_m$, respectively.  

Denoting the concentrations of $D$ and $T$ in the cytosol by $c_{D}^{}$ and $c_{T}^{}$ and by $m_{D}^{}$ and $m_{T}^{}$ on the membrane, respectively, the reaction-diffusion equations read 
\begin{subequations}	
\begin{align}
	\partial_t c_{T}^{}
	&= \, D_c \, \Delta \, c_{T}^{} 
	 + \lambda \, c_{D}^{} 
	 \, , \\
	\partial_t c_{D}^{}
	&= \,  D_c \, \Delta \, c_{D}^{}
	 - \lambda \, c_{D}^{} 
	 \, , \\
	\partial_t m_{T}^{}
	&= \,  D_m \, \Delta_m \, m_{T}^{} 
	 + (k_+ \, \tilde c_{T}^{} - k_- \, m_{T}^{})
	 + \, k^{}_{mT} \, 
	   m_{T}^{} \, \tilde c_{T}^{} 
     \, , \\
	\partial_t m_{D}^{}
	&= \,  D_m \, \Delta_m \, m_{D}^{} \, 
	 + (k_+  \, \tilde c_{D}^{} \, - k_- \, m_{D}^{})
	 + k^{\phantom{}}_{mD} \,  
	   m_{D}^{}  \; \tilde c_{D}^{}
	  \, .
\end{align}
\label{eq:one_ntpase_MaRD}
\end{subequations}
As before, reactive and diffusive fluxes balance at the membrane-cytosol boundary
\begin{subequations}
\begin{align}
	D_c \, \nabla_{\perp} c_{T}^{}{\mid_m}
	&=  - (k_+  \, + \, k^{}_{mT} \, 
	   m_{T}^{}) \, \tilde c_{T}^{} 
	\, \\
	D_c \, \nabla_{\perp} c_{D}^{}{\mid_m} 
	&=  - (k_+  \, + \, k^{}_{mD} \, 
	   m_{D}^{}) \, \tilde c_{D}^{}	    + k_- \, (m_{D}^{} +  m_{T}^{}) 
	\, .
\end{align}
\label{eq:one_ntpase_bc}
\end{subequations}

\begin{widetext}

\begin{figure}[tb]
\centering
\includegraphics[width=1.02\linewidth]{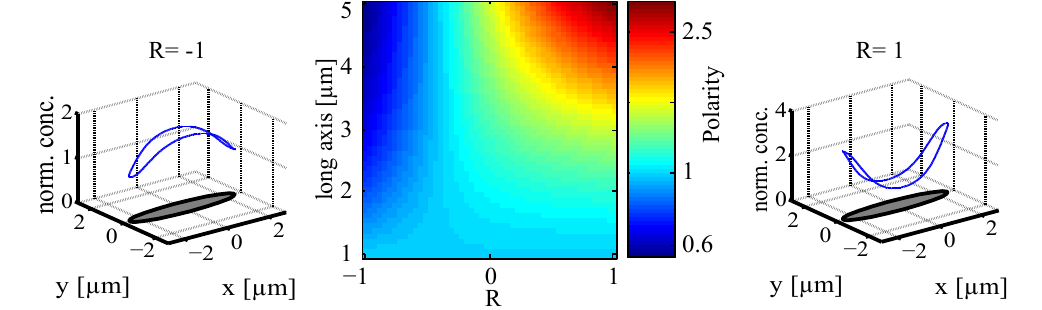}
\caption{
Membrane-bound proteins accumulate either at  midcell (left) or form a bipolar pattern with high protein densities at the cell poles (right). The left and right plots show the normalised concentration of the membrane density (blue curve) and the corresponding geometry of the cell (grey ellipse). 
The membrane density of the protein is divided by its minimum concentration (left: $113 \mu$m$^{-1}$, right: $100 \mu$m$^{-1}$) such that the minimum of the normalised density is $1$. 
The polarity ${\cal P} = m_\text{pole} / m_\text{midcell}$ (colour bar in plot is logarithmically spaced) of the pattern strongly depends on cell geometry and preference ${\cal R} = ( k^{}_{mD} - k^{}_{mT} ) / ( k^{}_{mD} + k^{}_{mT} )$ for the recruitment of a certain nucleotide state (middle);
the length of the short axis is fixed at $l =1 \,\mu$m, and we have used $k^{}_{mD}{+}k^{}_{mT} = 0.1 \, \mu$m/s.  While for large ${\cal R}$ (preferential recruitment of $D$) the proteins form a bipolar pattern on the membrane, the membrane-bound proteins accumulate at midcell for small $R$ (preferential recruitment of $T$). If the recruitment processes are balanced (${\cal R} =0$) the pattern is flat and polarity vanishes. The cell geometry determines how pronounced a pattern becomes: The more elongated the ellipse, the more sharply defined the pattern, which vanishes completely when the ellipse becomes a circle.
Reprinted from Ref.~\cite{Thalmeier_etal:2016} with permission form PNAS.
}
\label{fig:one_ntpase_polarity}      
\end{figure}

\end{widetext}

Solving this set of equations numerically in elliptical geometry reveals a series of striking features (Fig.\ref{fig:one_ntpase_polarity}): (i) In elongated cells the protein density on the membrane and in the cytosol is \textit{always} inhomogeneous, and reflects the local cell geometry. (ii) There are two distinct types of patterns: membrane-bound proteins either accumulate at midcell or form a bipolar pattern with high densities at both cell poles. (iii) The protein gradients scale with the size of the cell, i.e.\/ fully adapt to the geometry of the cell.

The type of polarity of these patterns is quantified by the ratio of the density of membrane-bound proteins located at the cell poles to that at midcell: ${\cal P} = m_\text{pole}/m_\text{midcell}$.
Accumulation occurs either at the cell pole or at midcell depending on the value of the preferential recruitment parameter ${\cal R} = (k^{\phantom{}}_{mD}{-}k^{}_{mT})/(k^{\phantom{}}_{mD}{+}k^{}_{mT})$: 
One finds that proteins accumulate at the cell poles (${\cal P}  > 1$) if there is a preference for cooperative binding of $D$ (${\cal R} > 0$).
Moreover, the polarity ${\cal P}$ of this bipolar pattern becomes more pronounced with increasing ${\cal R}$. 
In contrast, when cooperative binding favours $T$ (${\cal R} < 0$), proteins accumulate at midcell (${\cal P} < 1$). 
Thus, the sign of the recruitment preference ${\cal R}$ for a protein in a particular nucleotide state  controls the type, while its magnitude determines the amplitude of the pattern.  
With increasing eccentricity of the ellipse, the respective pattern becomes more sharply defined; for a spherical geometry the pattern vanishes. 
In summary, cell geometry controls the definition of the pattern, and the preference for membrane recruitment of a certain nucleotide state determines both the location on the cell membrane where the proteins accumulate and how pronounced this accumulation becomes.

\begin{figure}[!t]
\includegraphics[width=1.0\linewidth]{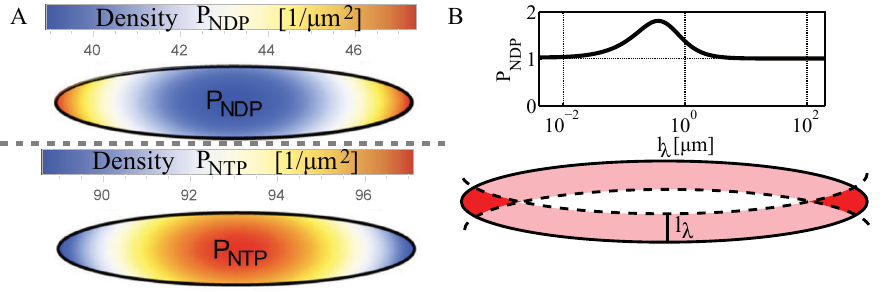}
\caption{ \textbf{Membrane affinity controls, and recruitment amplifies adaptation to geometry.} The cells used for the numerical studies have a length of $L = 5 \, \mu$m and a width of $l = 1 \, \mu$m.
\textbf{A,} Even when recruitment is turned off, $T$ and $D$ form inhomogeneous density profiles in the cytosol. $D$ accumulates close to the poles and is depleted at mid-cell. In contrast, $T$ exhibits high concentration at mid-cell and a low concentration at the poles. The attachment and detachment rates are set to $1 \, \mu$m/s and $1$s$^{-1}$,  respectively, which gives a penetration depth $\ell_\lambda \approx 1.6\, \mu$m. 
\textbf{B,} Illustration of the source-degradation mechanism for the spatial segregation of cytosolic $D$ and $T$. All proteins that detach from the membrane are in an NDP-bound state and can undergo nucleotide exchange, the range of $D$ in the cytosol is limited to a penetration depth $\ell_\lambda$ (dashed lines); here $\ell_\lambda = 0.35 \, \mu$m. At the poles this reaction volume receives input from opposing faces of the membrane resulting in an accumulation of cytosolic $D$ (dark red). The magnitude of this accumulation  depends on the penetration depth. The polarity ${\cal P}_\text{NDP} = m^\text{pole}_d/m^\text{mid-cell}_d$ of membrane-bound $D$ plotted as a function of $\ell_\lambda$ shows a maximum at $\ell_\lambda \approx 0.35 \, \mu$m and vanishes in the limits of large as well as small penetration depths.
Reprinted from Ref.~\cite{Thalmeier_etal:2016} with permission form PNAS.
}
\label{fig:one_ntpase_cytosolic_gradients}      
\end{figure}

What is the origin of these polar patterns and their features? To answer this question in the clearest possible way, it is instructive to consider the limiting case where positive feedback effects on recruitment are absent and the dynamics hence are fully linear. Then, Eqs.\ref{eq:one_ntpase_MaRD}-\ref{eq:one_ntpase_bc} imply that both the total concentration of proteins on the membrane, $m= m_{D}^{} + m_{T}^{}$, and in the cytosol, $c= c_{D}^{} + c_{T}^{}$, are spatially uniform if the detailed balance condition $k_+  \, \tilde c = k_- \, m$ holds for the exchange of proteins between the cytosol and the membrane.   
This uniformity in total protein density, however, does not imply uniformity in the densities of the active and inactive protein species, either on the cell membrane or in the cytosol! 
The origin of this effect is purely geometrical, and it is linked to the finite time required for nucleotide exchange in the cytosol.
Heuristically, this can be seen as follows (Fig. \ref{fig:one_ntpase_cytosolic_gradients}A). As only inactive proteins $D$ are released from the membrane they act as a source of cytosolic proteins. 
In the cytosol they are then reactivated through  nucleotide exchange, which is effectively equivalent to depleting the cytoplasmic compartment of inactive proteins.
This in turn implies the formation of a gradient of inactive proteins and a corresponding, oppositely oriented gradient of active proteins as one moves away from the membrane into the cytosol. 
As is known from standard source-degradation processes, the ensuing density profile for $D$ in the cytosol is exponential, with the decay length being set by $\ell_\lambda = \sqrt{D_c/\lambda}$.

Due to membrane curvature these reaction volumes overlap close to the cell poles (Fig. \ref{fig:one_ntpase_cytosolic_gradients}B, bottom), which implies an accumulation of $D$ at the cell poles. 
The effect becomes stronger with increasing membrane curvature. Moreover, there is an optimal value for the penetration depth $\ell_\lambda$, roughly equal to a third of the length $l$ of the short cell axis, that maximises accumulation of $D$ at the cell poles (Fig.\ref{fig:one_ntpase_cytosolic_gradients}B, top). As $\ell_\lambda$ becomes larger than $l$, the effect weakens, because the reaction volumes from opposite membrane sites also overlap at mid-cell. In the limit where $\ell_\lambda$ is much smaller than the membrane curvature at the poles, the overlap vanishes, and with it the accumulation of $D$ at the poles. 
More generally, these heuristic arguments imply that the local ratio of the reaction volume for nucleotide exchange to the available membrane surface is the factor that explains the dependence of the protein distribution on cell geometry. 

\section{\textit{In vitro} reconstitution and theoretical analysis of Min protein pattern formation}

A key step towards understanding pattern-formation mechanisms in biological systems is the identification of the essential functional modules that facilitate the formation of certain patterns. 
In living systems, such an identification is strongly impeded by the vast amount of potentially interacting and, therefore, interdependent components.
A common strategy for tackling the complexity of biological systems is mathematical modelling, which has been discussed in the previous section of this chapter. 
While mathematical analysis is able to identify possible mechanisms of pattern formation, it is also based on a priori assumptions about the biological system under consideration. 
However, these assumptions need to be tested by suitable experiments.
Ideally, a conclusive comparison between theory and experiment requires the ability to isolate the essential players of the pattern forming dynamics and reconstitute them in a minimal system lacking any other potential interactions and allowing for precise control of parameters, such as protein concentrations or geometric boundaries. 

\begin{figure}[t]
\includegraphics[width=\linewidth]{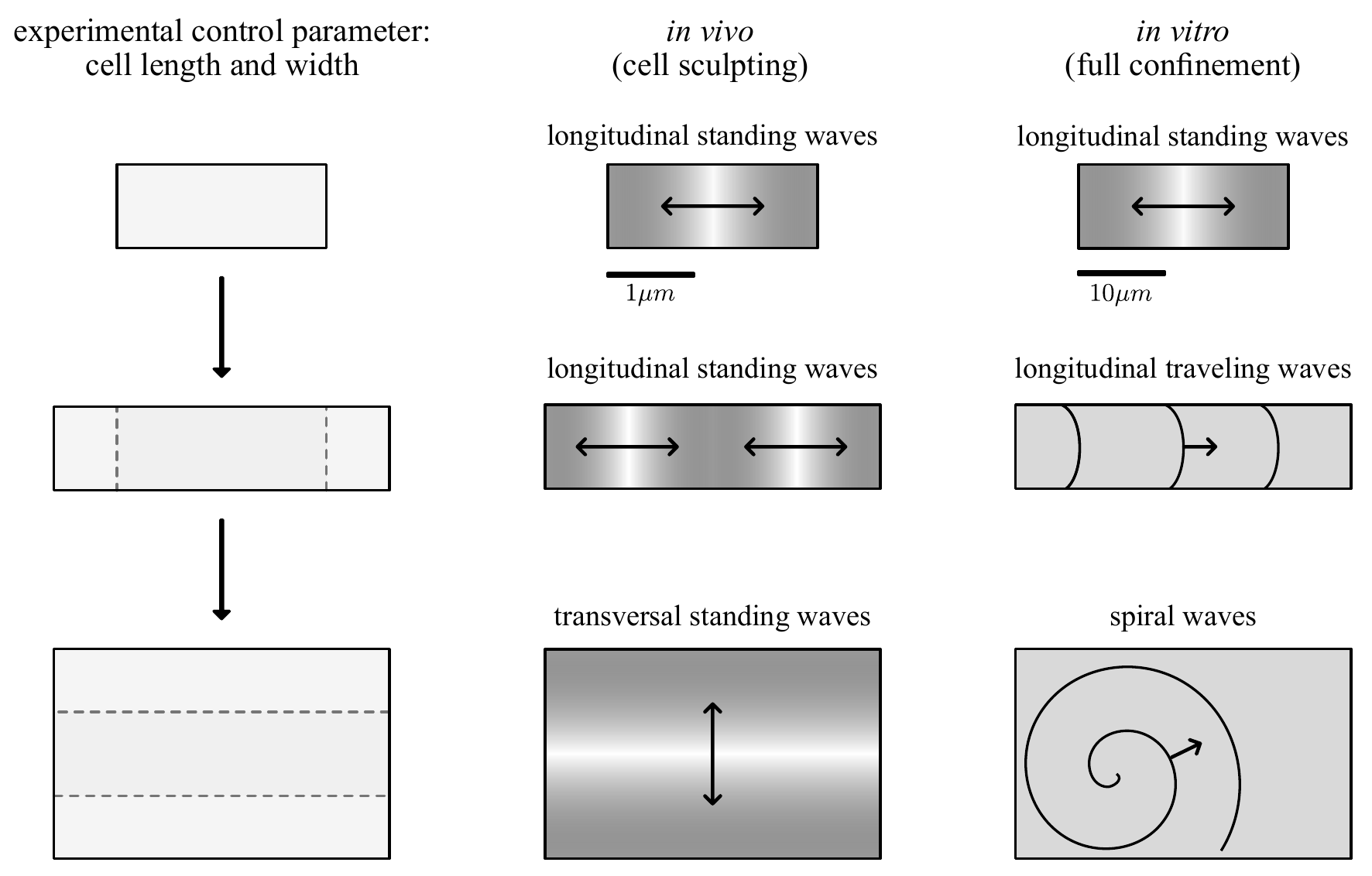}
\caption{\textbf{Min protein patterns \textit{in vivo} vs \textit{in vitro}}. Schematic depiction of the phenomenology observed in experiments when the system geometry is changed. For small systems the patterns in reconstituted systems \cite{Caspi_Dekker:2016} are similar to intracellular dynamics \cite{Wu_etal:2015}, showing pole-to-pole oscillations (with different length scales) in both cases. However, as the system length and width are increased, patterns appear that are not normally seen \textit{in vivo}.}
\label{fig:invivo_vs_invitro_pattern}      
\end{figure}

A major breakthrough in this regard was the successful \textit{in vitro} reconstitution of Min protein patterns in a lipid bilayer assay \cite{Loose_etal:2008}.
These experiments demonstrated that a flat lipid bilayer surface coupled to a cytosolic solution containing only MinD, MinE, and ATP is sufficient for the formation of membrane bound Min protein patterns. 
However, the patterns observed in reconstituted (\textit{in vitro}) systems significantly differed from the intracellular patterns found \textit{in vivo} (Fig.~\ref{fig:invivo_vs_invitro_pattern}).
While the majority of patterns found \textit{in vivo} can be viewed as standing waves with a wavelength matching the cell length, the patterns on the flat membrane are travelling and spiral waves with wavelengths one order of magnitude greater than the typical length of \textit{E. coli}. 

\subsection{A kaleidoscope of \textit{in vitro} patterns}

\begin{widetext}

\begin{figure}[t]
\includegraphics[width=1.05\textwidth]{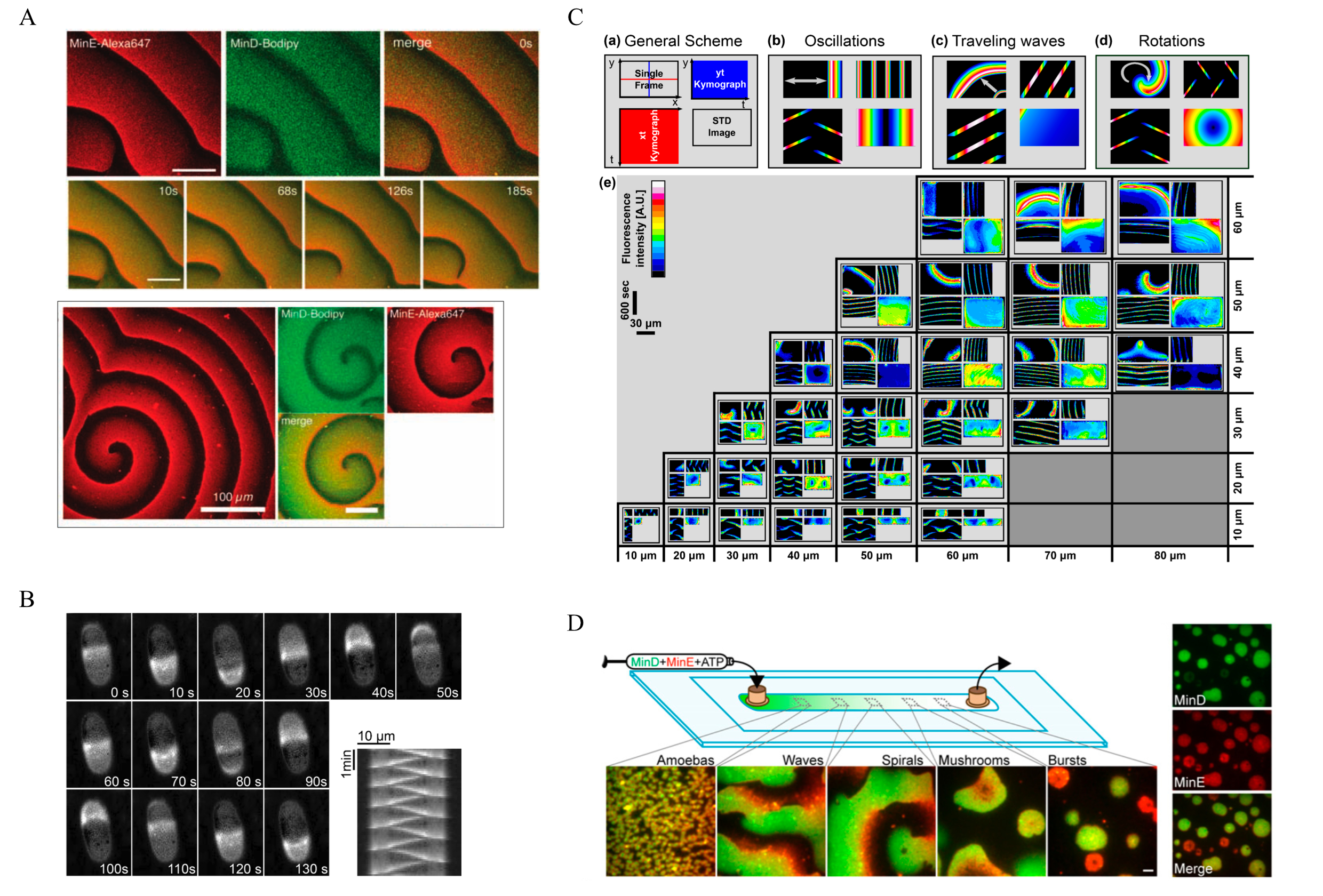}
\caption{
\textbf{Min patterns \textit{in vitro}.} 
\textbf{A,} Spiral- and travelling-wave patterns observed on flat lipid bilayers. From Ref.~\cite{Loose_etal:2008}. Reprinted with permission from AAAS. 
\textbf{B,} Pole-to-pole oscillations in semi-confined PDMS grooves. Reprinted with permission from Ref.~\cite{Zieske_Schwille:2013}, copyright 2013 Wiley-VCH Verlag GmbH and Co. KGaA, Weinheim, Germany.
\textbf{C,} Standing waves, travelling waves, and spiral waves observed in fully confined microfluidic chambers with different lateral dimensions. Adapted from Ref.~\protect \cite{Caspi_Dekker:2016} under the CC BY 4.0 license.
\textbf{D,} Exotic Min protein patterns on flat lipid bilayers in large laterally extended flow cells showing different phenomenology depending on the distance to the outlet and inlet of the flow cell device. Reprinted from Ref.~\protect\cite{Vecchiarelli_etal:2016} with permission form PNAS.
}
\label{fig:in_vitro_pattern}      
\end{figure}

\end{widetext}

The successful reconstitution of Min protein patterns on flat lipid bilayers stimulated a plethora of \textit{in vitro} experiments that studied Min protein dynamics under various circumstances and revealed a true kaleidoscope of patterns (Fig.~\ref{fig:in_vitro_pattern}).
On flat lipid bilayers one observed spiral and travelling wave patterns, and a varying degree of spatial coherence sometimes verging on chemical turbulence \cite{Loose_etal:2011}.
Other experiments constrained the Min protein dynamics geometrically to small membrane patches \cite{Schweizer_etal:2012}, semi-open PDMS grooves with varying lipid composition \cite{Zieske_Schwille:2013},  lipid-interfaced droplets \cite{Zieske_etal:2016}, and bilayer coated three-dimensional chambers of various shapes and sizes \cite{Caspi_Dekker:2016}. 
Strikingly, the observed patterns show a very broad range of characteristics and varying degrees of sensitivity to the geometry of the enclosing membrane.
Other experiments were performed in large, laterally extended flow cell devices with a flat lipid bilayer of varying lipid composition attached at the bottom \cite{Ivanov_Mizuuchi:2010}.
These experiments showed that Min protein patterns are formed even when there is hydrodynamic flow in the cytosol. 
Furthermore, these experiments revealed the capability of Min protein dynamics to form exotic patterns sharing characteristics of travelling waves and stationary patterns alike \cite{Ivanov_Mizuuchi:2010}. 

Despite these intensive experimental efforts, a quantitative reconstitution of Min protein patterns observed \textit{in vivo} has not been achieved. 
Instead a broad range of different patterns has been found, all of which exhibit wavelengths that are several times larger than that of the \textit{in vivo} pattern.
The pole-to-pole patterns that are observed in (semi-)confined compartments \cite{Zieske_Schwille:2014, Caspi_Dekker:2016} most closely resemble those seen \textit{in vivo}. 
Interestingly, this resemblance is limited to geometries with dimensions below the typical wavelength of the pattern. 
In these systems the characteristic pole-to-pole oscillation is observed \textit{in vivo} as well as \textit{in vitro}. 
If the length and width of the confined system are increased, the reconstituted \textit{in vitro} experiments \cite{Caspi_Dekker:2016} predominantly show traveling and spiral wave patterns, whereas \textit{in vivo} experiments show longitudinal and transversal standing waves \cite{Wu_etal:2015, Wu_etal:2016}. 
This suggests that the underlying mechanisms (dynamic instabilities) are actually not the same\footnote{We note that travelling wave patterns have also been observed \textit{in vivo} \cite{Bonny_etal:2013}, albeit only upon massive over-expression of MinD and MinE, leading to highly elevated intracellular protein densities and pathological phenomenology \cite{Sliusarenko_etal:2011} relative to the wild type. While the exact protein densities in the experiments have not been measured, this observation is consistent with the observation of travelling waves in fully confined compartments, where the protein densities inside microfluidic chambers were also elevated \cite{Caspi_Dekker:2016}. 
For further discussion of the effect of protein densities we refer the reader to section \protect\ref{sec:polychotomy}.}. 
While longitudinal and transversal standing waves have also been observed in semi-confined PDMS grooves of specific sizes \cite{Zieske_Schwille:2014}, the patterns became chaotic in these experiments when the system size was increased \cite{Zieske_Schwille:2014}.

Given these ambiguous results, how can we reconcile the kaleidoscope of \textit{in vitro} patterns and the range of \textit{in vivo} patterns?
In the following, we discuss how theory can shed some light on these bewildering results.
As we will see, a key problem with the interpretation of recent \textit{in vitro} reconstitution experiments and their comparison to \textit{in vivo} dynamics lies in the lack of the ceteris paribus condition, i.e.\/ conditions where only one control parameter is varied while the rest are held constant. 
Achieving quantitative control over all parameters will be the key goal for future experiments.

\subsection{The polychotomy of Min protein patterns} 
\label{sec:polychotomy}

All experimental evidence supports the assumption that the Min system can be understood as a reaction-diffusion system driven by nonlinear (cooperative) protein interactions. 
Therefore, we can expect that Min protein dynamics will share generic features of such nonlinear systems. 
In particular, as is well known in the field of nonlinear dynamics, even very simple models can produce a broad variety of patterns~\cite{Gray_Scott:1983, Gray_Scott:1984, Gray_Scott:1985, Pearson:1993, Lee_etal:1993}. 
Moreover, which patterns are observed depends on the parameters of the system. 
In the classical mathematical theory these parameters are the coefficients of the (non-)linear interactions (representing the ``kinetics''), as well as the diffusion coefficients.
 
Diffusion coefficients (in the cytosol) have been measured \textit{in vivo} \cite{Meacci_etal:2006} and \textit{in vitro} \cite{Loose_etal:2008, Loose_etal:2011}, and they can be controlled experimentally by the addition of crowding agents \cite{Schweizer_etal:2012, Caspi_Dekker:2016}. 
Kinetic parameters of the Min system are much more difficult to measure and to control. 
However, diffusion coefficients and kinetic rates are not the only control parameters. 
Most of the classical literature in nonlinear dynamics neither accounts for system geometry nor for the mass-conserving nature of bio-molecular interactions. 
This might explain why the fact that system geometry as well as protein densities can be key control parameters of the system's dynamics is often overlooked. 
The effect of changes in these parameters is not necessarily restricted to changes in the length- and time-scales of the dynamics (e.g. wavelength, wave speed, and oscillation period), but can also induce qualitative changes and transitions between patterns.

One clear difference between the reconstituted Min system on flat lipid bilayers and the intracellular system in \textit{E. coli} is the vastly increased ratio of cytosolic volume to membrane surface in the \textit{in vitro} system, where the height of the system is of the order of milimetres, instead of $\mu$m, in the living system. 
A recent theoretical analysis \cite{halatek_frey:2018} has shown that increasing this volume--to--surface ratio leads to an increased wavelength of the pattern. 
This prediction agrees with the experimental observation of a reduced wavelength of the Min protein patterns in fully confined geometries \cite{Caspi_Dekker:2016} that mimic the \textit{in vivo} membrane-to-cytosol ratio more closely than does the flat lipid bilayer.
Strikingly, even when cytosolic diffusion was reduced to \textit{in vivo} levels, these experiments still showed a $3$- to $4$-fold increased wavelength in confined compartments compared to the intracellular patterns -- emphasising an apparent dichotomy between patterns observed \textit{in vivo} and \textit{in vitro}. 

However, the surface--to--volume ratio is not the only difference between the intracellular and the reconstituted Min systems. Another is the particle number or effective density of MinD as well as MinE.
At first glance there is no apparent difference between the protein concentrations \textit{in vivo} and \textit{in vitro}, since the concentrations in all reconstituted systems are adjusted to the intracellular concentrations which are about $1 \, \mu$M for MinD and MinE. 
However, it is important to note that these are the average cytosolic densities with no proteins attached to the membrane. 
Since all cytosolic proteins are able to bind to the membrane\footnote{Either directly, or by complex formation as for MinDE complexes.}, the total number of cytosolic proteins determines the upper bound for the maximal membrane densities. 
Hence, even if the average cytosolic densities in the reconstituted system are identical to typical intracellular concentrations, the crucial control parameter is the ratio of cytosolic volume to membrane surface.
\textit{In vivo}, a cytosolic density of about $1 \mu$M yields a number of proteins that can easily be absorbed by the membrane and still remain up to two orders of magnitude below the saturation limit.\footnote{Assuming a cylindrical geometry for simplicity, the volume to surface ratio is $\sim r/2$, i.e.\/ well below $1 \,\mu$m for typical cell radii $r$.}
However, in the reconstituted system with flat lipid bilayer the volume to surface ratio is given by the bulk height $h$. 
For the typical bulk height on the order of millimetres, less than 1\% of all proteins can bind to the membrane before saturation due to volume exclusion is reached.
As a consequence, the protein densities at and on the membrane are highly increased in the reconstituted system compared to the situation \textit{in vivo}, despite the average cytosolic densities being identical.
Note that the densities of membrane-bound proteins are directly involved in the recruitment process which represents the only intrinsically nonlinear interaction in the Min system (cf.\/ section \ref{sec:MaRD_Min}). 
As such, one can expect that changes in the average protein densities on the membrane affect the system dynamics in a significant way.
Indeed, estimates of the concentration on the flat lipid bilayer show that the density across a wave profile is about two orders of magnitude higher than the typical protein densities on the intracellular membrane \cite{Loose_etal:2011}.
The same can be assumed to be the case for reconstituted Min oscillations in semi-open PDMS grooves \cite{Zieske_Schwille:2013, Zieske_Schwille:2014}, since the dynamics are initialised with a high cytosolic column above the grooves which is only removed after the onset of pattern formation (and therefore membrane accumulation). 
Elevated protein densities were also found for the reconstituted Min patterns in confined chambers \cite{Caspi_Dekker:2016} since these are based on a microfluidic device. 
As proteins accumulate on the membrane while the flow is still active, the density at the inlet is merely a lower bound for the actual protein densities in the individual chambers. 
Measurements of the protein fluorescence inside the confined chambers after careful calibration show that the total densities of MinD and MinE and the MinE/MinD ratios are increased and are broadly distributed \cite{Caspi_Dekker:2016}. 
A similar result can be expected for Min protein dynamics in large, laterally extended flow cells where diverse wave patterns are observed \cite{Ivanov_Mizuuchi:2010, Vecchiarelli_etal:2016}.

To put these findings from the \textit{in vitro} reconstruction of Min protein pattern in the context of the theoretical framework, the broad variation of volume to surface ratios, total protein numbers, and MinE/MinD density ratios, is a crucial aspect to consider (cf. \cite{Halatek_Frey:2014}). 
The theoretical analysis of the skeleton model, Eqs.\ref{eq:RD_cytosol}--\ref{eq:RD_boundary}, has shown that all these quantities are key control parameters for the system dynamics. 
An increase in any of these values (total density, density ratio, volume/surface ratio) can lead to a Turing- or Hopf-instability \cite{halatek_frey:2018}. 
In the latter case,  each point on the membrane can be considered to be an individual chemical oscillator, and the laterally extended system a field of diffusively coupled oscillators \cite{halatek_frey:2018}.
Such dynamics describe a broad class of systems well documented in the classic nonlinear dynamics literature \cite{Aranson_Kramer:2002}. 
Key characteristics of oscillatory media are spiral and travelling patterns, as well as various manifestations of chemical turbulence. 
All these phenomena can be observed in the reconstituted Min system \cite{denk_etal:2018}.
From this point of view, the observed dichotomy rather appears as a polychotomy, not only between \textit{in vivo} and \textit{in vitro}, but between the many different experimental setups. 
Its origin lies in the broad distribution of control parameters and emphasises the diversity of Min protein dynamics on a phenomenological and mechanistic level.

\section{Discussion and outlook}

As outlined in this chapter, the recent focus on the quantitative study of pattern formation in biological systems has led to conceptually new approaches in theory and experiments.
Among the important milestones are the inclusion of cell geometry and an explicit distinction between cell membrane and cytosolic volume in theoretical models, as well as the identification of particle numbers and cell geometry as major control parameters of the self-organisation processes that lead to pattern formation. 
While these efforts enabled the quantitative study of biological pattern formation within the theoretical framework of nonlinear dynamics, experimental advances in  \textit{in vitro} reconstitution opened new ways to probe, study, and design protein pattern formation as well as controlled minimal systems.
Due to its simplicity, the \textit{E. coli} Min system has been the subject of intensive theoretical and experimental investigation, establishing it as a paradigm for protein pattern formation. 
In contrast, the eukaryotic systems discussed here remain far less well understood. 
In part, this is due to a higher degree of complexity and redundancy in these systems. 
For example, PAR networks involve several different molecular players in the anterior and posterior PAR components respectively, and also interact with dynamic cytoskeletal structures and physical triggers \cite{Goehring:2014}.
Accordingly, the \textit{in vitro} reconstitution of eukaryotic pattern-forming systems is typically more challenging compared to bacterial systems. 
Yet, efforts to experimentally reconstitute even basic aspects of such pattern-forming systems \textit{in vitro} could substantially enhance our understanding of their underlying mechanisms via control and perturbation of the experimental conditions. 

For the Min system, several key questions remain to be answered. Central is the experimental control over system parameters that gives rise to the multitude of observed patterns. 
Future research may reveal additional chemical states of MinD as well as MinE or additional chemical reactions that refine the hitherto identified skeleton network.
While this will affect the number of chemical components and reaction terms one has to take into account in the mathematical model, it does not change the overall structure of the set of reaction-diffusion equations: 
(1) Fast cytosolic diffusion is coupled to slow membrane dynamics by chemical reactions that conserve protein number. 
(2) Nucleotide exchange in the cytosol implies that active MinD is spatially separated from the reactive membrane. As a consequence, the cytosol serves as a repository for active MinD. 
(3) MinD and MinE remain the only conserved species. The sum of individual components of each species, regardless of the number of components, will always be a conserved quantity.

Open questions relating to molecular details of Min protein interaction concern the roles of membrane binding and conformational state switching of MinE \cite{Park_etal:2011}. 
Only a combined approach, in which the theoretical model is constrained and supported by unambiguous experimental data, has the potential to truly relate molecular ``design'' features of Min proteins to defined roles in pattern formation.

In summary, protein pattern formation plays key roles in many essential biological processes from bacteria to animals, including cell polarisation and division. 
Combined theoretical and experimental approaches have established important principles of pattern-forming protein systems. 
Perhaps the most crucial feature that has emerged from these research efforts is the identification of the cytosol as a depot. This depot enables the system to store proteins and redistribute them throughout the system. 
Cytosolic diffusion is the key process that detects the local shape of the membrane, and it is this explicit dependence on geometry that is imprinted on membrane-bound protein patterns. 

\acknowledgements
We thank Fridtjof Brauns, Yaron Caspi, Cees Dekker, Jonas Denk, and Fabai Wu for helpful discussions.
This research was supported by the German Excellence Initiative via the program ``NanoSystems Initiative Munich'' (NIM),  and the Deutsche Forschungsgemeinschaft (DFG) via project A09 and B02 within the Collaborative Research Center (SFB 1032) ``Nanoagents for spatio-temporal control of molecular and cellular reactions''. 
SK is supported by a DFG fellowship through QBM.


\begin{thebibliography}{97}%
\makeatletter
\providecommand \@ifxundefined [1]{%
 \@ifx{#1\undefined}
}%
\providecommand \@ifnum [1]{%
 \ifnum #1\expandafter \@firstoftwo
 \else \expandafter \@secondoftwo
 \fi
}%
\providecommand \@ifx [1]{%
 \ifx #1\expandafter \@firstoftwo
 \else \expandafter \@secondoftwo
 \fi
}%
\providecommand \natexlab [1]{#1}%
\providecommand \enquote  [1]{``#1''}%
\providecommand \bibnamefont  [1]{#1}%
\providecommand \bibfnamefont [1]{#1}%
\providecommand \citenamefont [1]{#1}%
\providecommand \href@noop [0]{\@secondoftwo}%
\providecommand \href [0]{\begingroup \@sanitize@url \@href}%
\providecommand \@href[1]{\@@startlink{#1}\@@href}%
\providecommand \@@href[1]{\endgroup#1\@@endlink}%
\providecommand \@sanitize@url [0]{\catcode `\\12\catcode `\$12\catcode
  `\&12\catcode `\#12\catcode `\^12\catcode `\_12\catcode `\%12\relax}%
\providecommand \@@startlink[1]{}%
\providecommand \@@endlink[0]{}%
\providecommand \url  [0]{\begingroup\@sanitize@url \@url }%
\providecommand \@url [1]{\endgroup\@href {#1}{\urlprefix }}%
\providecommand \urlprefix  [0]{URL }%
\providecommand \Eprint [0]{\href }%
\providecommand \doibase [0]{http://dx.doi.org/}%
\providecommand \selectlanguage [0]{\@gobble}%
\providecommand \bibinfo  [0]{\@secondoftwo}%
\providecommand \bibfield  [0]{\@secondoftwo}%
\providecommand \translation [1]{[#1]}%
\providecommand \BibitemOpen [0]{}%
\providecommand \bibitemStop [0]{}%
\providecommand \bibitemNoStop [0]{.\EOS\space}%
\providecommand \EOS [0]{\spacefactor3000\relax}%
\providecommand \BibitemShut  [1]{\csname bibitem#1\endcsname}%
\let\auto@bib@innerbib\@empty
\bibitem [{\citenamefont {Thom}(1983)}]{Thom:1983}%
  \BibitemOpen
  \bibfield  {author} {\bibinfo {author} {\bibfnamefont {R.}~\bibnamefont
  {Thom}},\ }\href
  {http://books.google.de/books?id=erETAQAAIAAJ&dq=inauthor:Thom+intitle:Morphogenesis&hl=&cd=1&source=gbs_api}
  {{\selectlanguage {English}\emph {\bibinfo {title} {{Mathematical models of
  morphogenesis}}}}}\ (\bibinfo {year} {1983})\BibitemShut {NoStop}%
\bibitem [{\citenamefont {Turing}(1952)}]{Turing:1952}%
  \BibitemOpen
  \bibfield  {author} {\bibinfo {author} {\bibfnamefont {A.~M.}\ \bibnamefont
  {Turing}},\ }\href {\doibase 10.1007/BF02459572} {\bibfield  {journal}
  {\bibinfo  {journal} {Philosophical Transactions of the Royal Society of
  London. Series B, Biological Sciences}\ }\textbf {\bibinfo {volume} {237}},\
  \bibinfo {pages} {37} (\bibinfo {year} {1952})}\BibitemShut {NoStop}%
\bibitem [{\citenamefont {Guckenheimer}\ and\ \citenamefont
  {Holmes}(2013)}]{Guckenheimer:2013}%
  \BibitemOpen
  \bibfield  {author} {\bibinfo {author} {\bibfnamefont {J.}~\bibnamefont
  {Guckenheimer}}\ and\ \bibinfo {author} {\bibfnamefont {P.~J.}\ \bibnamefont
  {Holmes}},\ }\href
  {http://books.google.de/books?id=XYIpBAAAQBAJ&pg=PA222&dq=inauthor:guckenheimer+intitle:Nonlinear+dynamics&hl=&cd=1&source=gbs_api}
  {\emph {\bibinfo {title} {{Nonlinear Oscillations, Dynamical Systems, and
  Bifurcations of Vector Fields}}}}\ (\bibinfo  {publisher} {Springer Science
  {\&} Business Media},\ \bibinfo {year} {2013})\BibitemShut {NoStop}%
\bibitem [{\citenamefont {Cross}\ and\ \citenamefont
  {Greenside}(2009)}]{Cross_Greenside:Book}%
  \BibitemOpen
  \bibfield  {author} {\bibinfo {author} {\bibfnamefont {M.}~\bibnamefont
  {Cross}}\ and\ \bibinfo {author} {\bibfnamefont {H.}~\bibnamefont
  {Greenside}},\ }\href@noop {} {\emph {\bibinfo {title} {Pattern Formation and
  Dynamics in Nonequilibrium Systems}}}\ (\bibinfo  {publisher} {Cambridge
  University Press},\ \bibinfo {year} {2009})\BibitemShut {NoStop}%
\bibitem [{\citenamefont {Wedlich-Soldner}\ \emph {et~al.}(2003)\citenamefont
  {Wedlich-Soldner}, \citenamefont {Altschuler}, \citenamefont {Wu},\ and\
  \citenamefont {Li}}]{Wedlich-Soldner_etal:2003}%
  \BibitemOpen
  \bibfield  {author} {\bibinfo {author} {\bibfnamefont {R.}~\bibnamefont
  {Wedlich-Soldner}}, \bibinfo {author} {\bibfnamefont {S.}~\bibnamefont
  {Altschuler}}, \bibinfo {author} {\bibfnamefont {L.}~\bibnamefont {Wu}}, \
  and\ \bibinfo {author} {\bibfnamefont {R.}~\bibnamefont {Li}},\ }\href@noop
  {} {\bibfield  {journal} {\bibinfo  {journal} {Science}\ }\textbf {\bibinfo
  {volume} {299}},\ \bibinfo {pages} {1231} (\bibinfo {year}
  {2003})}\BibitemShut {NoStop}%
\bibitem [{\citenamefont {Florian}\ and\ \citenamefont
  {Geiger}(2010)}]{Florian_Geiger:2010}%
  \BibitemOpen
  \bibfield  {author} {\bibinfo {author} {\bibfnamefont {M.~C.}\ \bibnamefont
  {Florian}}\ and\ \bibinfo {author} {\bibfnamefont {H.}~\bibnamefont
  {Geiger}},\ }\href@noop {} {\bibfield  {journal} {\bibinfo  {journal} {Stem
  Cells}\ }\textbf {\bibinfo {volume} {28}},\ \bibinfo {pages} {1623} (\bibinfo
  {year} {2010})}\BibitemShut {NoStop}%
\bibitem [{\citenamefont {Molendijk}\ \emph {et~al.}(2001)\citenamefont
  {Molendijk}, \citenamefont {Bischoff}, \citenamefont {Rajendrakumar},
  \citenamefont {Friml}, \citenamefont {Braun}, \citenamefont {Gilroy},\ and\
  \citenamefont {Palme}}]{Molendijk_etal:2001}%
  \BibitemOpen
  \bibfield  {author} {\bibinfo {author} {\bibfnamefont {A.~J.}\ \bibnamefont
  {Molendijk}}, \bibinfo {author} {\bibfnamefont {F.}~\bibnamefont {Bischoff}},
  \bibinfo {author} {\bibfnamefont {C.~S.}\ \bibnamefont {Rajendrakumar}},
  \bibinfo {author} {\bibfnamefont {J.}~\bibnamefont {Friml}}, \bibinfo
  {author} {\bibfnamefont {M.}~\bibnamefont {Braun}}, \bibinfo {author}
  {\bibfnamefont {S.}~\bibnamefont {Gilroy}}, \ and\ \bibinfo {author}
  {\bibfnamefont {K.}~\bibnamefont {Palme}},\ }\href@noop {} {\bibfield
  {journal} {\bibinfo  {journal} {EMBO J}\ }\textbf {\bibinfo {volume} {20}},\
  \bibinfo {pages} {2779} (\bibinfo {year} {2001})}\BibitemShut {NoStop}%
\bibitem [{\citenamefont {Gu}\ \emph {et~al.}(2003)\citenamefont {Gu},
  \citenamefont {Vernoud}, \citenamefont {Fu},\ and\ \citenamefont
  {Yang}}]{Gu_etal:2003}%
  \BibitemOpen
  \bibfield  {author} {\bibinfo {author} {\bibfnamefont {Y.}~\bibnamefont
  {Gu}}, \bibinfo {author} {\bibfnamefont {V.}~\bibnamefont {Vernoud}},
  \bibinfo {author} {\bibfnamefont {Y.}~\bibnamefont {Fu}}, \ and\ \bibinfo
  {author} {\bibfnamefont {Z.}~\bibnamefont {Yang}},\ }\href@noop {} {\bibfield
   {journal} {\bibinfo  {journal} {J Exp Bot}\ }\textbf {\bibinfo {volume}
  {54}},\ \bibinfo {pages} {93} (\bibinfo {year} {2003})}\BibitemShut {NoStop}%
\bibitem [{\citenamefont {Goehring}\ \emph {et~al.}(2011)\citenamefont
  {Goehring}, \citenamefont {Trong}, \citenamefont {Bois}, \citenamefont
  {Chowdhury}, \citenamefont {Nicola}, \citenamefont {Hyman},\ and\
  \citenamefont {Grill}}]{Goehring_etal:2011}%
  \BibitemOpen
  \bibfield  {author} {\bibinfo {author} {\bibfnamefont {N.~W.}\ \bibnamefont
  {Goehring}}, \bibinfo {author} {\bibfnamefont {P.~K.}\ \bibnamefont {Trong}},
  \bibinfo {author} {\bibfnamefont {J.~S.}\ \bibnamefont {Bois}}, \bibinfo
  {author} {\bibfnamefont {D.}~\bibnamefont {Chowdhury}}, \bibinfo {author}
  {\bibfnamefont {E.~M.}\ \bibnamefont {Nicola}}, \bibinfo {author}
  {\bibfnamefont {A.~A.}\ \bibnamefont {Hyman}}, \ and\ \bibinfo {author}
  {\bibfnamefont {S.~W.}\ \bibnamefont {Grill}},\ }\href@noop {} {\bibfield
  {journal} {\bibinfo  {journal} {Science (New York, NY)}\ }\textbf {\bibinfo
  {volume} {334}},\ \bibinfo {pages} {1137} (\bibinfo {year}
  {2011})}\BibitemShut {NoStop}%
\bibitem [{\citenamefont {Bement}\ \emph {et~al.}(2015)\citenamefont {Bement},
  \citenamefont {Leda}, \citenamefont {Moe}, \citenamefont {Kita},
  \citenamefont {Larson}, \citenamefont {Golding}, \citenamefont {Pfeuti},
  \citenamefont {Su}, \citenamefont {Miller}, \citenamefont {Goryachev},\ and\
  \citenamefont {von Dassow}}]{Bement_etal:2015}%
  \BibitemOpen
  \bibfield  {author} {\bibinfo {author} {\bibfnamefont {W.~M.}\ \bibnamefont
  {Bement}}, \bibinfo {author} {\bibfnamefont {M.}~\bibnamefont {Leda}},
  \bibinfo {author} {\bibfnamefont {A.~M.}\ \bibnamefont {Moe}}, \bibinfo
  {author} {\bibfnamefont {A.~M.}\ \bibnamefont {Kita}}, \bibinfo {author}
  {\bibfnamefont {M.~E.}\ \bibnamefont {Larson}}, \bibinfo {author}
  {\bibfnamefont {A.~E.}\ \bibnamefont {Golding}}, \bibinfo {author}
  {\bibfnamefont {C.}~\bibnamefont {Pfeuti}}, \bibinfo {author} {\bibfnamefont
  {K.-C.}\ \bibnamefont {Su}}, \bibinfo {author} {\bibfnamefont {A.~L.}\
  \bibnamefont {Miller}}, \bibinfo {author} {\bibfnamefont {A.~B.}\
  \bibnamefont {Goryachev}}, \ and\ \bibinfo {author} {\bibfnamefont
  {G.}~\bibnamefont {von Dassow}},\ }\href@noop {} {\bibfield  {journal}
  {\bibinfo  {journal} {Nature Cell Biology}\ }\textbf {\bibinfo {volume}
  {17}},\ \bibinfo {pages} {1471} (\bibinfo {year} {2015})}\BibitemShut
  {NoStop}%
\bibitem [{\citenamefont {Desai}\ and\ \citenamefont
  {Mitchison}(2003)}]{Desai_Mitchison:2003}%
  \BibitemOpen
  \bibfield  {author} {\bibinfo {author} {\bibfnamefont {A.}~\bibnamefont
  {Desai}}\ and\ \bibinfo {author} {\bibfnamefont {T.~J.}\ \bibnamefont
  {Mitchison}},\ }\href@noop {} {\bibfield  {journal} {\bibinfo  {journal}
  {dx.doi.org.emedien.ub.uni-muenchen.de}\ } (\bibinfo {year}
  {2003})}\BibitemShut {NoStop}%
\bibitem [{\citenamefont {Varga}\ \emph {et~al.}(2006)\citenamefont {Varga},
  \citenamefont {Helenius}, \citenamefont {Tanaka}, \citenamefont {Hyman},
  \citenamefont {Tanaka},\ and\ \citenamefont {Howard}}]{Varga2006}%
  \BibitemOpen
  \bibfield  {author} {\bibinfo {author} {\bibfnamefont {V.}~\bibnamefont
  {Varga}}, \bibinfo {author} {\bibfnamefont {J.}~\bibnamefont {Helenius}},
  \bibinfo {author} {\bibfnamefont {K.}~\bibnamefont {Tanaka}}, \bibinfo
  {author} {\bibfnamefont {A.~A.}\ \bibnamefont {Hyman}}, \bibinfo {author}
  {\bibfnamefont {T.~U.}\ \bibnamefont {Tanaka}}, \ and\ \bibinfo {author}
  {\bibfnamefont {J.}~\bibnamefont {Howard}},\ }\href {\doibase
  10.1038/ncb1462} {\bibfield  {journal} {\bibinfo  {journal} {Nat. Cell
  Biol.}\ }\textbf {\bibinfo {volume} {8}},\ \bibinfo {pages} {957} (\bibinfo
  {year} {2006})}\BibitemShut {NoStop}%
\bibitem [{\citenamefont {Varga}\ \emph {et~al.}(2009)\citenamefont {Varga},
  \citenamefont {Leduc}, \citenamefont {Bormuth}, \citenamefont {Diez},\ and\
  \citenamefont {Howard}}]{Varga2009}%
  \BibitemOpen
  \bibfield  {author} {\bibinfo {author} {\bibfnamefont {V.}~\bibnamefont
  {Varga}}, \bibinfo {author} {\bibfnamefont {C.}~\bibnamefont {Leduc}},
  \bibinfo {author} {\bibfnamefont {V.}~\bibnamefont {Bormuth}}, \bibinfo
  {author} {\bibfnamefont {S.}~\bibnamefont {Diez}}, \ and\ \bibinfo {author}
  {\bibfnamefont {J.}~\bibnamefont {Howard}},\ }\href {\doibase
  10.1016/j.cell.2009.07.032} {\bibfield  {journal} {\bibinfo  {journal}
  {Cell}\ }\textbf {\bibinfo {volume} {138}},\ \bibinfo {pages} {1174}
  (\bibinfo {year} {2009})}\BibitemShut {NoStop}%
\bibitem [{\citenamefont {Reese}\ \emph {et~al.}(2011)\citenamefont {Reese},
  \citenamefont {Melbinger},\ and\ \citenamefont {Frey}}]{Reese_etal:2011}%
  \BibitemOpen
  \bibfield  {author} {\bibinfo {author} {\bibfnamefont {L.}~\bibnamefont
  {Reese}}, \bibinfo {author} {\bibfnamefont {A.}~\bibnamefont {Melbinger}}, \
  and\ \bibinfo {author} {\bibfnamefont {E.}~\bibnamefont {Frey}},\ }\href@noop
  {} {\bibfield  {journal} {\bibinfo  {journal} {Biophysical journal}\ }\textbf
  {\bibinfo {volume} {101}},\ \bibinfo {pages} {2190} (\bibinfo {year}
  {2011})}\BibitemShut {NoStop}%
\bibitem [{\citenamefont {Melbinger}\ \emph {et~al.}(2012)\citenamefont
  {Melbinger}, \citenamefont {Reese},\ and\ \citenamefont
  {Frey}}]{Melbinger_etal:2012}%
  \BibitemOpen
  \bibfield  {author} {\bibinfo {author} {\bibfnamefont {A.}~\bibnamefont
  {Melbinger}}, \bibinfo {author} {\bibfnamefont {L.}~\bibnamefont {Reese}}, \
  and\ \bibinfo {author} {\bibfnamefont {E.}~\bibnamefont {Frey}},\ }\href@noop
  {} {\bibfield  {journal} {\bibinfo  {journal} {Physical review letters}\
  }\textbf {\bibinfo {volume} {108}},\ \bibinfo {pages} {258104} (\bibinfo
  {year} {2012})}\BibitemShut {NoStop}%
\bibitem [{\citenamefont {Reese}\ \emph {et~al.}(2014)\citenamefont {Reese},
  \citenamefont {Melbinger},\ and\ \citenamefont {Frey}}]{Reese2014}%
  \BibitemOpen
  \bibfield  {author} {\bibinfo {author} {\bibfnamefont {L.}~\bibnamefont
  {Reese}}, \bibinfo {author} {\bibfnamefont {A.}~\bibnamefont {Melbinger}}, \
  and\ \bibinfo {author} {\bibfnamefont {E.}~\bibnamefont {Frey}},\ }\href
  {\doibase 10.1098/rsfs.2014.0031} {\bibfield  {journal} {\bibinfo  {journal}
  {Interface Focus}\ }\textbf {\bibinfo {volume} {4}} (\bibinfo {year}
  {2014}),\ 10.1098/rsfs.2014.0031}\BibitemShut {NoStop}%
\bibitem [{\citenamefont {Raskin}\ and\ \citenamefont
  {de~Boer}(1999{\natexlab{a}})}]{Raskin_deBoer:1999a}%
  \BibitemOpen
  \bibfield  {author} {\bibinfo {author} {\bibfnamefont {D.~M.}\ \bibnamefont
  {Raskin}}\ and\ \bibinfo {author} {\bibfnamefont {P.~A.}\ \bibnamefont
  {de~Boer}},\ }\href
  {http://eutils.ncbi.nlm.nih.gov/entrez/eutils/elink.fcgi?dbfrom=pubmed&id=10515933&retmode=ref&cmd=prlinks}
  {\bibfield  {journal} {\bibinfo  {journal} {Journal of Bacteriology}\
  }\textbf {\bibinfo {volume} {181}},\ \bibinfo {pages} {6419} (\bibinfo {year}
  {1999}{\natexlab{a}})}\BibitemShut {NoStop}%
\bibitem [{\citenamefont {Raskin}\ and\ \citenamefont
  {de~Boer}(1999{\natexlab{b}})}]{Raskin_deBoer:1999b}%
  \BibitemOpen
  \bibfield  {author} {\bibinfo {author} {\bibfnamefont {D.~M.}\ \bibnamefont
  {Raskin}}\ and\ \bibinfo {author} {\bibfnamefont {P.~A.}\ \bibnamefont
  {de~Boer}},\ }\href
  {http://eutils.ncbi.nlm.nih.gov/entrez/eutils/elink.fcgi?dbfrom=pubmed&id=10220403&retmode=ref&cmd=prlinks}
  {\bibfield  {journal} {\bibinfo  {journal} {Proceedings of the National
  Academy of Sciences}\ }\textbf {\bibinfo {volume} {96}},\ \bibinfo {pages}
  {4971} (\bibinfo {year} {1999}{\natexlab{b}})}\BibitemShut {NoStop}%
\bibitem [{\citenamefont {Hu}\ and\ \citenamefont
  {Lutkenhaus}(1999)}]{Hu_Lutkenhaus:1999}%
  \BibitemOpen
  \bibfield  {author} {\bibinfo {author} {\bibfnamefont {Z.}~\bibnamefont
  {Hu}}\ and\ \bibinfo {author} {\bibfnamefont {J.}~\bibnamefont
  {Lutkenhaus}},\ }\href@noop {} {\bibfield  {journal} {\bibinfo  {journal}
  {Molecular Microbiology}\ }\textbf {\bibinfo {volume} {34}},\ \bibinfo
  {pages} {82} (\bibinfo {year} {1999})}\BibitemShut {NoStop}%
\bibitem [{\citenamefont {Lutkenhaus}(2007)}]{Lutkenhaus:2007}%
  \BibitemOpen
  \bibfield  {author} {\bibinfo {author} {\bibfnamefont {J.}~\bibnamefont
  {Lutkenhaus}},\ }\href {\doibase 10.1146/annurev.biochem.75.103004.142652}
  {\bibfield  {journal} {\bibinfo  {journal} {Annual review of biochemistry}\
  }\textbf {\bibinfo {volume} {76}},\ \bibinfo {pages} {539} (\bibinfo {year}
  {2007})}\BibitemShut {NoStop}%
\bibitem [{\citenamefont {Szeto}\ \emph {et~al.}(2002)\citenamefont {Szeto},
  \citenamefont {Rowland}, \citenamefont {Rothfield},\ and\ \citenamefont
  {King}}]{Szeto_etal:2002}%
  \BibitemOpen
  \bibfield  {author} {\bibinfo {author} {\bibfnamefont {T.~H.}\ \bibnamefont
  {Szeto}}, \bibinfo {author} {\bibfnamefont {S.~L.}\ \bibnamefont {Rowland}},
  \bibinfo {author} {\bibfnamefont {L.~I.}\ \bibnamefont {Rothfield}}, \ and\
  \bibinfo {author} {\bibfnamefont {G.~F.}\ \bibnamefont {King}},\ }\href
  {\doibase 10.1073/pnas.232590599} {\bibfield  {journal} {\bibinfo  {journal}
  {Proceedings of the National Academy of Sciences}\ }\textbf {\bibinfo
  {volume} {99}},\ \bibinfo {pages} {15693} (\bibinfo {year}
  {2002})}\BibitemShut {NoStop}%
\bibitem [{\citenamefont {Hu}\ and\ \citenamefont
  {Lutkenhaus}(2003)}]{Hu_Lutkenhaus:2003}%
  \BibitemOpen
  \bibfield  {author} {\bibinfo {author} {\bibfnamefont {Z.}~\bibnamefont
  {Hu}}\ and\ \bibinfo {author} {\bibfnamefont {J.}~\bibnamefont
  {Lutkenhaus}},\ }\href@noop {} {\bibfield  {journal} {\bibinfo  {journal}
  {Molecular Microbiology}\ }\textbf {\bibinfo {volume} {47}},\ \bibinfo
  {pages} {345} (\bibinfo {year} {2003})}\BibitemShut {NoStop}%
\bibitem [{\citenamefont {Lackner}\ \emph {et~al.}(2003)\citenamefont
  {Lackner}, \citenamefont {Raskin},\ and\ \citenamefont
  {de~Boer}}]{Lackner_etal:2003}%
  \BibitemOpen
  \bibfield  {author} {\bibinfo {author} {\bibfnamefont {L.~L.}\ \bibnamefont
  {Lackner}}, \bibinfo {author} {\bibfnamefont {D.~M.}\ \bibnamefont {Raskin}},
  \ and\ \bibinfo {author} {\bibfnamefont {P.~A.~J.}\ \bibnamefont {de~Boer}},\
  }\href {\doibase 10.1128/JB.185.3.735-749.2003} {\bibfield  {journal}
  {\bibinfo  {journal} {Journal of Bacteriology}\ }\textbf {\bibinfo {volume}
  {185}},\ \bibinfo {pages} {735} (\bibinfo {year} {2003})}\BibitemShut
  {NoStop}%
\bibitem [{\citenamefont {Mileykovskaya}\ \emph {et~al.}(2003)\citenamefont
  {Mileykovskaya}, \citenamefont {Fishov}, \citenamefont {Fu}, \citenamefont
  {Corbin}, \citenamefont {Margolin},\ and\ \citenamefont
  {Dowhan}}]{Mileykovskaya_etal:2003}%
  \BibitemOpen
  \bibfield  {author} {\bibinfo {author} {\bibfnamefont {E.}~\bibnamefont
  {Mileykovskaya}}, \bibinfo {author} {\bibfnamefont {I.}~\bibnamefont
  {Fishov}}, \bibinfo {author} {\bibfnamefont {X.}~\bibnamefont {Fu}}, \bibinfo
  {author} {\bibfnamefont {B.~D.}\ \bibnamefont {Corbin}}, \bibinfo {author}
  {\bibfnamefont {W.}~\bibnamefont {Margolin}}, \ and\ \bibinfo {author}
  {\bibfnamefont {W.}~\bibnamefont {Dowhan}},\ }\href {\doibase
  10.1074/jbc.M302603200} {\bibfield  {journal} {\bibinfo  {journal} {Journal
  of Biological Chemistry}\ }\textbf {\bibinfo {volume} {278}},\ \bibinfo
  {pages} {22193} (\bibinfo {year} {2003})}\BibitemShut {NoStop}%
\bibitem [{\citenamefont {Hu}\ \emph {et~al.}(1999)\citenamefont {Hu},
  \citenamefont {Mukherjee}, \citenamefont {Pichoff},\ and\ \citenamefont
  {Lutkenhaus}}]{Hu_etal:1999}%
  \BibitemOpen
  \bibfield  {author} {\bibinfo {author} {\bibfnamefont {Z.}~\bibnamefont
  {Hu}}, \bibinfo {author} {\bibfnamefont {A.}~\bibnamefont {Mukherjee}},
  \bibinfo {author} {\bibfnamefont {S.}~\bibnamefont {Pichoff}}, \ and\
  \bibinfo {author} {\bibfnamefont {J.}~\bibnamefont {Lutkenhaus}},\
  }\href@noop {} {\bibfield  {journal} {\bibinfo  {journal} {Proceedings of the
  National Academy of Sciences}\ }\textbf {\bibinfo {volume} {96}},\ \bibinfo
  {pages} {14819} (\bibinfo {year} {1999})}\BibitemShut {NoStop}%
\bibitem [{\citenamefont {Hu}\ and\ \citenamefont
  {Lutkenhaus}(2001)}]{Hu_Lutkenhaus:2001}%
  \BibitemOpen
  \bibfield  {author} {\bibinfo {author} {\bibfnamefont {Z.}~\bibnamefont
  {Hu}}\ and\ \bibinfo {author} {\bibfnamefont {J.}~\bibnamefont
  {Lutkenhaus}},\ }\href@noop {} {\bibfield  {journal} {\bibinfo  {journal}
  {Molecular Cell}\ }\textbf {\bibinfo {volume} {7}},\ \bibinfo {pages} {1337}
  (\bibinfo {year} {2001})}\BibitemShut {NoStop}%
\bibitem [{\citenamefont {Hu}\ \emph {et~al.}(2002)\citenamefont {Hu},
  \citenamefont {Gogol},\ and\ \citenamefont {Lutkenhaus}}]{Hu_etal:2002}%
  \BibitemOpen
  \bibfield  {author} {\bibinfo {author} {\bibfnamefont {Z.}~\bibnamefont
  {Hu}}, \bibinfo {author} {\bibfnamefont {E.~P.}\ \bibnamefont {Gogol}}, \
  and\ \bibinfo {author} {\bibfnamefont {J.}~\bibnamefont {Lutkenhaus}},\
  }\href@noop {} {\bibfield  {journal} {\bibinfo  {journal} {Proceedings of the
  National Academy of Sciences}\ }\textbf {\bibinfo {volume} {99}},\ \bibinfo
  {pages} {6761} (\bibinfo {year} {2002})}\BibitemShut {NoStop}%
\bibitem [{\citenamefont {Park}\ \emph {et~al.}(2011)\citenamefont {Park},
  \citenamefont {Wu}, \citenamefont {Battaile}, \citenamefont {Lovell},
  \citenamefont {Holyoak},\ and\ \citenamefont {Lutkenhaus}}]{Park_etal:2011}%
  \BibitemOpen
  \bibfield  {author} {\bibinfo {author} {\bibfnamefont {K.-T.}\ \bibnamefont
  {Park}}, \bibinfo {author} {\bibfnamefont {W.}~\bibnamefont {Wu}}, \bibinfo
  {author} {\bibfnamefont {K.~P.}\ \bibnamefont {Battaile}}, \bibinfo {author}
  {\bibfnamefont {S.}~\bibnamefont {Lovell}}, \bibinfo {author} {\bibfnamefont
  {T.}~\bibnamefont {Holyoak}}, \ and\ \bibinfo {author} {\bibfnamefont
  {J.}~\bibnamefont {Lutkenhaus}},\ }\href@noop {} {\bibfield  {journal}
  {\bibinfo  {journal} {Cell}\ }\textbf {\bibinfo {volume} {146}},\ \bibinfo
  {pages} {396} (\bibinfo {year} {2011})}\BibitemShut {NoStop}%
\bibitem [{\citenamefont {Shih}\ \emph {et~al.}(2011)\citenamefont {Shih},
  \citenamefont {Huang}, \citenamefont {Lai}, \citenamefont {Liao},
  \citenamefont {Lee}, \citenamefont {Chang}, \citenamefont {Mak},
  \citenamefont {Hsieh},\ and\ \citenamefont {Lin}}]{Shih_etal:2011}%
  \BibitemOpen
  \bibfield  {author} {\bibinfo {author} {\bibfnamefont {Y.-L.}\ \bibnamefont
  {Shih}}, \bibinfo {author} {\bibfnamefont {K.-F.}\ \bibnamefont {Huang}},
  \bibinfo {author} {\bibfnamefont {H.-M.}\ \bibnamefont {Lai}}, \bibinfo
  {author} {\bibfnamefont {J.-H.}\ \bibnamefont {Liao}}, \bibinfo {author}
  {\bibfnamefont {C.-S.}\ \bibnamefont {Lee}}, \bibinfo {author} {\bibfnamefont
  {C.-M.}\ \bibnamefont {Chang}}, \bibinfo {author} {\bibfnamefont {H.-M.}\
  \bibnamefont {Mak}}, \bibinfo {author} {\bibfnamefont {C.-W.}\ \bibnamefont
  {Hsieh}}, \ and\ \bibinfo {author} {\bibfnamefont {C.-C.}\ \bibnamefont
  {Lin}},\ }\href {\doibase 10.1371/journal.pone.0021425} {\bibfield  {journal}
  {\bibinfo  {journal} {PLoS ONE}\ }\textbf {\bibinfo {volume} {6}},\ \bibinfo
  {pages} {e21425} (\bibinfo {year} {2011})}\BibitemShut {NoStop}%
\bibitem [{\citenamefont {Loose}\ \emph
  {et~al.}(2011{\natexlab{a}})\citenamefont {Loose}, \citenamefont {Kruse},\
  and\ \citenamefont {Schwille}}]{Loose_etal:2011_review}%
  \BibitemOpen
  \bibfield  {author} {\bibinfo {author} {\bibfnamefont {M.}~\bibnamefont
  {Loose}}, \bibinfo {author} {\bibfnamefont {K.}~\bibnamefont {Kruse}}, \ and\
  \bibinfo {author} {\bibfnamefont {P.}~\bibnamefont {Schwille}},\ }\href
  {\doibase 10.1146/annurev-biophys-042910-155332} {\bibfield  {journal}
  {\bibinfo  {journal} {Annual Review Of Biophysics}\ }\textbf {\bibinfo
  {volume} {40}},\ \bibinfo {pages} {315} (\bibinfo {year}
  {2011}{\natexlab{a}})}\BibitemShut {NoStop}%
\bibitem [{\citenamefont {Lutkenhaus}(2012)}]{Lutkenhaus:2012}%
  \BibitemOpen
  \bibfield  {author} {\bibinfo {author} {\bibfnamefont {J.}~\bibnamefont
  {Lutkenhaus}},\ }\href {\doibase 10.1016/j.tim.2012.05.002} {\bibfield
  {journal} {\bibinfo  {journal} {Trends In Microbiology}\ }\textbf {\bibinfo
  {volume} {20}},\ \bibinfo {pages} {411} (\bibinfo {year} {2012})}\BibitemShut
  {NoStop}%
\bibitem [{\citenamefont {Loose}\ and\ \citenamefont
  {Mitchison}(2014)}]{Loose_Mitchison:2014}%
  \BibitemOpen
  \bibfield  {author} {\bibinfo {author} {\bibfnamefont {M.}~\bibnamefont
  {Loose}}\ and\ \bibinfo {author} {\bibfnamefont {T.~J.}\ \bibnamefont
  {Mitchison}},\ }\href@noop {} {\bibfield  {journal} {\bibinfo  {journal}
  {Nature Cell Biology}\ }\textbf {\bibinfo {volume} {16}},\ \bibinfo {pages}
  {38} (\bibinfo {year} {2014})}\BibitemShut {NoStop}%
\bibitem [{\citenamefont {Denk}\ \emph {et~al.}(2016)\citenamefont {Denk},
  \citenamefont {Huber}, \citenamefont {Reithmann},\ and\ \citenamefont
  {Frey}}]{Denk_etal:2016}%
  \BibitemOpen
  \bibfield  {author} {\bibinfo {author} {\bibfnamefont {J.}~\bibnamefont
  {Denk}}, \bibinfo {author} {\bibfnamefont {L.}~\bibnamefont {Huber}},
  \bibinfo {author} {\bibfnamefont {E.}~\bibnamefont {Reithmann}}, \ and\
  \bibinfo {author} {\bibfnamefont {E.}~\bibnamefont {Frey}},\ }\href@noop {}
  {\bibfield  {journal} {\bibinfo  {journal} {Physical Review Letters}\
  }\textbf {\bibinfo {volume} {116}},\ \bibinfo {pages} {178301} (\bibinfo
  {year} {2016})}\BibitemShut {NoStop}%
\bibitem [{\citenamefont {Ramirez}\ \emph {et~al.}(2016)\citenamefont
  {Ramirez}, \citenamefont {Garcia-Soriano}, \citenamefont {Raso},
  \citenamefont {Feingold}, \citenamefont {Rivas},\ and\ \citenamefont
  {Schwille}}]{Ramirez_etal:2016}%
  \BibitemOpen
  \bibfield  {author} {\bibinfo {author} {\bibfnamefont {D.}~\bibnamefont
  {Ramirez}}, \bibinfo {author} {\bibfnamefont {D.~A.}\ \bibnamefont
  {Garcia-Soriano}}, \bibinfo {author} {\bibfnamefont {A.}~\bibnamefont
  {Raso}}, \bibinfo {author} {\bibfnamefont {M.}~\bibnamefont {Feingold}},
  \bibinfo {author} {\bibfnamefont {G.}~\bibnamefont {Rivas}}, \ and\ \bibinfo
  {author} {\bibfnamefont {P.}~\bibnamefont {Schwille}},\ }\href@noop {}
  {\bibfield  {journal} {\bibinfo  {journal} {biorxiv}\ } (\bibinfo {year}
  {2016})}\BibitemShut {NoStop}%
\bibitem [{\citenamefont {Bi}\ and\ \citenamefont {Park}(2012)}]{Bi_Park:2012}%
  \BibitemOpen
  \bibfield  {author} {\bibinfo {author} {\bibfnamefont {E.}~\bibnamefont
  {Bi}}\ and\ \bibinfo {author} {\bibfnamefont {H.-O.}\ \bibnamefont {Park}},\
  }\href {\doibase 10.1534/genetics.111.132886} {\bibfield  {journal} {\bibinfo
   {journal} {Genetics}\ }\textbf {\bibinfo {volume} {191}},\ \bibinfo {pages}
  {347} (\bibinfo {year} {2012})}\BibitemShut {NoStop}%
\bibitem [{\citenamefont {Freisinger}\ \emph {et~al.}(2013)\citenamefont
  {Freisinger}, \citenamefont {Kl{\"u}nder}, \citenamefont {Johnson},
  \citenamefont {M{\"u}ller}, \citenamefont {Pichler}, \citenamefont {Beck},
  \citenamefont {Costanzo}, \citenamefont {Boone}, \citenamefont {Cerione},
  \citenamefont {Frey},\ and\ \citenamefont
  {Wedlich-Soldner}}]{Freisinger_etal:2013}%
  \BibitemOpen
  \bibfield  {author} {\bibinfo {author} {\bibfnamefont {T.}~\bibnamefont
  {Freisinger}}, \bibinfo {author} {\bibfnamefont {B.}~\bibnamefont
  {Kl{\"u}nder}}, \bibinfo {author} {\bibfnamefont {J.}~\bibnamefont
  {Johnson}}, \bibinfo {author} {\bibfnamefont {N.}~\bibnamefont {M{\"u}ller}},
  \bibinfo {author} {\bibfnamefont {G.}~\bibnamefont {Pichler}}, \bibinfo
  {author} {\bibfnamefont {G.}~\bibnamefont {Beck}}, \bibinfo {author}
  {\bibfnamefont {M.}~\bibnamefont {Costanzo}}, \bibinfo {author}
  {\bibfnamefont {C.}~\bibnamefont {Boone}}, \bibinfo {author} {\bibfnamefont
  {R.~A.}\ \bibnamefont {Cerione}}, \bibinfo {author} {\bibfnamefont
  {E.}~\bibnamefont {Frey}}, \ and\ \bibinfo {author} {\bibfnamefont
  {R.}~\bibnamefont {Wedlich-Soldner}},\ }\href {\doibase 10.1038/ncomms2795}
  {\bibfield  {journal} {\bibinfo  {journal} {Nature communications}\ }\textbf
  {\bibinfo {volume} {4}},\ \bibinfo {pages} {1807} (\bibinfo {year}
  {2013})}\BibitemShut {NoStop}%
\bibitem [{\citenamefont {Goryachev}\ and\ \citenamefont
  {Pokhilko}(2008)}]{Goryachev_Pokhilko:2008}%
  \BibitemOpen
  \bibfield  {author} {\bibinfo {author} {\bibfnamefont {A.~B.}\ \bibnamefont
  {Goryachev}}\ and\ \bibinfo {author} {\bibfnamefont {A.~V.}\ \bibnamefont
  {Pokhilko}},\ }\href@noop {} {\bibfield  {journal} {\bibinfo  {journal} {FEBS
  Lett.}\ }\textbf {\bibinfo {volume} {582}},\ \bibinfo {pages} {1437}
  (\bibinfo {year} {2008})}\BibitemShut {NoStop}%
\bibitem [{\citenamefont {Kl{\"u}nder}\ \emph {et~al.}(2013)\citenamefont
  {Kl{\"u}nder}, \citenamefont {Freisinger}, \citenamefont {Wedlich-Soldner},\
  and\ \citenamefont {Frey}}]{Klunder_etal:2013}%
  \BibitemOpen
  \bibfield  {author} {\bibinfo {author} {\bibfnamefont {B.}~\bibnamefont
  {Kl{\"u}nder}}, \bibinfo {author} {\bibfnamefont {T.}~\bibnamefont
  {Freisinger}}, \bibinfo {author} {\bibfnamefont {R.}~\bibnamefont
  {Wedlich-Soldner}}, \ and\ \bibinfo {author} {\bibfnamefont {E.}~\bibnamefont
  {Frey}},\ }\href {\doibase 10.1371/journal.pcbi.1003396} {\bibfield
  {journal} {\bibinfo  {journal} {Plos Computational Biology}\ }\textbf
  {\bibinfo {volume} {9}},\ \bibinfo {pages} {e1003396} (\bibinfo {year}
  {2013})}\BibitemShut {NoStop}%
\bibitem [{\citenamefont {Bose}\ \emph {et~al.}(2001)\citenamefont {Bose},
  \citenamefont {Irazoqui}, \citenamefont {Moskow}, \citenamefont {Bardes},
  \citenamefont {Zyla},\ and\ \citenamefont {Lew}}]{Bose_etal:2001}%
  \BibitemOpen
  \bibfield  {author} {\bibinfo {author} {\bibfnamefont {I.}~\bibnamefont
  {Bose}}, \bibinfo {author} {\bibfnamefont {J.~E.}\ \bibnamefont {Irazoqui}},
  \bibinfo {author} {\bibfnamefont {J.~J.}\ \bibnamefont {Moskow}}, \bibinfo
  {author} {\bibfnamefont {E.~S.}\ \bibnamefont {Bardes}}, \bibinfo {author}
  {\bibfnamefont {T.~R.}\ \bibnamefont {Zyla}}, \ and\ \bibinfo {author}
  {\bibfnamefont {D.~J.}\ \bibnamefont {Lew}},\ }\href@noop {} {\bibfield
  {journal} {\bibinfo  {journal} {J. Biol. Chem.}\ }\textbf {\bibinfo {volume}
  {276}},\ \bibinfo {pages} {7176} (\bibinfo {year} {2001})}\BibitemShut
  {NoStop}%
\bibitem [{\citenamefont {Butty}\ \emph {et~al.}(2002)\citenamefont {Butty},
  \citenamefont {Perrinjaquet}, \citenamefont {Petit}, \citenamefont
  {Jaquenoud}, \citenamefont {Segall}, \citenamefont {Hofmann}, \citenamefont
  {Zwahlen},\ and\ \citenamefont {Peter}}]{Butty_etal:2002}%
  \BibitemOpen
  \bibfield  {author} {\bibinfo {author} {\bibfnamefont {A.-C.}\ \bibnamefont
  {Butty}}, \bibinfo {author} {\bibfnamefont {N.}~\bibnamefont {Perrinjaquet}},
  \bibinfo {author} {\bibfnamefont {A.}~\bibnamefont {Petit}}, \bibinfo
  {author} {\bibfnamefont {M.}~\bibnamefont {Jaquenoud}}, \bibinfo {author}
  {\bibfnamefont {J.~E.}\ \bibnamefont {Segall}}, \bibinfo {author}
  {\bibfnamefont {K.}~\bibnamefont {Hofmann}}, \bibinfo {author} {\bibfnamefont
  {C.}~\bibnamefont {Zwahlen}}, \ and\ \bibinfo {author} {\bibfnamefont
  {M.}~\bibnamefont {Peter}},\ }\href@noop {} {\bibfield  {journal} {\bibinfo
  {journal} {EMBO J.}\ }\textbf {\bibinfo {volume} {21}},\ \bibinfo {pages}
  {1565} (\bibinfo {year} {2002})}\BibitemShut {NoStop}%
\bibitem [{\citenamefont {Kozubowski}\ \emph {et~al.}(2008)\citenamefont
  {Kozubowski}, \citenamefont {Saito}, \citenamefont {Johnson}, \citenamefont
  {Howell}, \citenamefont {Zyla},\ and\ \citenamefont
  {Lew}}]{Kozubowski_etal:2008}%
  \BibitemOpen
  \bibfield  {author} {\bibinfo {author} {\bibfnamefont {L.}~\bibnamefont
  {Kozubowski}}, \bibinfo {author} {\bibfnamefont {K.}~\bibnamefont {Saito}},
  \bibinfo {author} {\bibfnamefont {J.~M.}\ \bibnamefont {Johnson}}, \bibinfo
  {author} {\bibfnamefont {A.~S.}\ \bibnamefont {Howell}}, \bibinfo {author}
  {\bibfnamefont {T.~R.}\ \bibnamefont {Zyla}}, \ and\ \bibinfo {author}
  {\bibfnamefont {D.~J.}\ \bibnamefont {Lew}},\ }\href@noop {} {\bibfield
  {journal} {\bibinfo  {journal} {Curr. Biol.}\ }\textbf {\bibinfo {volume}
  {18}},\ \bibinfo {pages} {1719} (\bibinfo {year} {2008})}\BibitemShut
  {NoStop}%
\bibitem [{\citenamefont {Bendezu}\ \emph {et~al.}(2015)\citenamefont
  {Bendezu}, \citenamefont {Vincenzetti}, \citenamefont {Vavylonis},
  \citenamefont {Wyss}, \citenamefont {Vogel},\ and\ \citenamefont
  {Martin}}]{Bendezu_etal:2015}%
  \BibitemOpen
  \bibfield  {author} {\bibinfo {author} {\bibfnamefont {F.~O.}\ \bibnamefont
  {Bendezu}}, \bibinfo {author} {\bibfnamefont {V.}~\bibnamefont
  {Vincenzetti}}, \bibinfo {author} {\bibfnamefont {D.}~\bibnamefont
  {Vavylonis}}, \bibinfo {author} {\bibfnamefont {R.}~\bibnamefont {Wyss}},
  \bibinfo {author} {\bibfnamefont {H.}~\bibnamefont {Vogel}}, \ and\ \bibinfo
  {author} {\bibfnamefont {S.~G.}\ \bibnamefont {Martin}},\ }\href@noop {}
  {\bibfield  {journal} {\bibinfo  {journal} {PLoS biology}\ }\textbf {\bibinfo
  {volume} {13}} (\bibinfo {year} {2015})}\BibitemShut {NoStop}%
\bibitem [{\citenamefont {Laan}\ \emph {et~al.}(2015)\citenamefont {Laan},
  \citenamefont {Koschwanez},\ and\ \citenamefont {Murray}}]{Laan_etal:2015}%
  \BibitemOpen
  \bibfield  {author} {\bibinfo {author} {\bibfnamefont {L.}~\bibnamefont
  {Laan}}, \bibinfo {author} {\bibfnamefont {J.~H.}\ \bibnamefont
  {Koschwanez}}, \ and\ \bibinfo {author} {\bibfnamefont {A.~W.}\ \bibnamefont
  {Murray}},\ }\href@noop {} {\bibfield  {journal} {\bibinfo  {journal}
  {eLife}\ }\textbf {\bibinfo {volume} {4}} (\bibinfo {year}
  {2015})}\BibitemShut {NoStop}%
\bibitem [{\citenamefont {Brauns}\ \emph {et~al.}()\citenamefont {Brauns},
  \citenamefont {Halatek}, \citenamefont {Laan},\ and\ \citenamefont
  {Frey}}]{Brauns_etal:unpubl}%
  \BibitemOpen
  \bibfield  {author} {\bibinfo {author} {\bibfnamefont {F.}~\bibnamefont
  {Brauns}}, \bibinfo {author} {\bibfnamefont {J.}~\bibnamefont {Halatek}},
  \bibinfo {author} {\bibfnamefont {L.}~\bibnamefont {Laan}}, \ and\ \bibinfo
  {author} {\bibfnamefont {E.}~\bibnamefont {Frey}},\ }\href@noop {}
  {}\BibitemShut {NoStop}%
\bibitem [{\citenamefont {Goldstein}\ and\ \citenamefont
  {Macara}(2007)}]{Goldstein_Macara:2007}%
  \BibitemOpen
  \bibfield  {author} {\bibinfo {author} {\bibfnamefont {B.}~\bibnamefont
  {Goldstein}}\ and\ \bibinfo {author} {\bibfnamefont {I.~G.}\ \bibnamefont
  {Macara}},\ }\href@noop {} {\bibfield  {journal} {\bibinfo  {journal}
  {Developmental Cell}\ }\textbf {\bibinfo {volume} {13}},\ \bibinfo {pages}
  {609} (\bibinfo {year} {2007})}\BibitemShut {NoStop}%
\bibitem [{\citenamefont {Munro}\ \emph {et~al.}(2004)\citenamefont {Munro},
  \citenamefont {Nance},\ and\ \citenamefont {Priess}}]{Munro_etal:2004}%
  \BibitemOpen
  \bibfield  {author} {\bibinfo {author} {\bibfnamefont {E.}~\bibnamefont
  {Munro}}, \bibinfo {author} {\bibfnamefont {J.}~\bibnamefont {Nance}}, \ and\
  \bibinfo {author} {\bibfnamefont {J.~R.}\ \bibnamefont {Priess}},\ }\href
  {\doibase 10.1016/j.devcel.2004.08.001} {\bibfield  {journal} {\bibinfo
  {journal} {Developmental Cell}\ }\textbf {\bibinfo {volume} {7}},\ \bibinfo
  {pages} {413} (\bibinfo {year} {2004})}\BibitemShut {NoStop}%
\bibitem [{\citenamefont {Goehring}(2014)}]{Goehring:2014}%
  \BibitemOpen
  \bibfield  {author} {\bibinfo {author} {\bibfnamefont {N.~W.}\ \bibnamefont
  {Goehring}},\ }\href@noop {} {\bibfield  {journal} {\bibinfo  {journal}
  {Experimental Cell Research}\ }\textbf {\bibinfo {volume} {328}},\ \bibinfo
  {pages} {258} (\bibinfo {year} {2014})}\BibitemShut {NoStop}%
\bibitem [{\citenamefont {Goehring}\ and\ \citenamefont
  {Grill}(2013)}]{Goehring_Grill:2013}%
  \BibitemOpen
  \bibfield  {author} {\bibinfo {author} {\bibfnamefont {N.~W.}\ \bibnamefont
  {Goehring}}\ and\ \bibinfo {author} {\bibfnamefont {S.~W.}\ \bibnamefont
  {Grill}},\ }\href {\doibase 10.1016/j.tcb.2012.10.009} {\bibfield  {journal}
  {\bibinfo  {journal} {Trends in cell biology}\ }\textbf {\bibinfo {volume}
  {23}},\ \bibinfo {pages} {72} (\bibinfo {year} {2013})}\BibitemShut {NoStop}%
\bibitem [{\citenamefont {Green}\ \emph {et~al.}(2012)\citenamefont {Green},
  \citenamefont {Paluch},\ and\ \citenamefont {Oegema}}]{Green_etal:2012}%
  \BibitemOpen
  \bibfield  {author} {\bibinfo {author} {\bibfnamefont {R.~A.}\ \bibnamefont
  {Green}}, \bibinfo {author} {\bibfnamefont {E.}~\bibnamefont {Paluch}}, \
  and\ \bibinfo {author} {\bibfnamefont {K.}~\bibnamefont {Oegema}},\
  }\href@noop {} {\bibfield  {journal} {\bibinfo  {journal} {Annual review of
  cell and developmental biology}\ }\textbf {\bibinfo {volume} {28}},\ \bibinfo
  {pages} {29} (\bibinfo {year} {2012})}\BibitemShut {NoStop}%
\bibitem [{\citenamefont {Gerdes}\ \emph {et~al.}(2010)\citenamefont {Gerdes},
  \citenamefont {Howard},\ and\ \citenamefont
  {Szardenings}}]{Gerdes_etal:2010}%
  \BibitemOpen
  \bibfield  {author} {\bibinfo {author} {\bibfnamefont {K.}~\bibnamefont
  {Gerdes}}, \bibinfo {author} {\bibfnamefont {M.}~\bibnamefont {Howard}}, \
  and\ \bibinfo {author} {\bibfnamefont {F.}~\bibnamefont {Szardenings}},\
  }\href {\doibase 10.1016/j.cell.2010.05.033} {\bibfield  {journal} {\bibinfo
  {journal} {Cell}\ }\textbf {\bibinfo {volume} {141}},\ \bibinfo {pages} {927}
  (\bibinfo {year} {2010})}\BibitemShut {NoStop}%
\bibitem [{\citenamefont {Bange}\ and\ \citenamefont
  {Sinning}(2013)}]{Bange_Sinning:2013}%
  \BibitemOpen
  \bibfield  {author} {\bibinfo {author} {\bibfnamefont {G.}~\bibnamefont
  {Bange}}\ and\ \bibinfo {author} {\bibfnamefont {I.}~\bibnamefont
  {Sinning}},\ }\href {\doibase 10.1038/nsmb.2605} {\bibfield  {journal}
  {\bibinfo  {journal} {Nature structural {\&} molecular biology}\ }\textbf
  {\bibinfo {volume} {20}},\ \bibinfo {pages} {776} (\bibinfo {year}
  {2013})}\BibitemShut {NoStop}%
\bibitem [{\citenamefont {Treuner-Lange}\ and\ \citenamefont
  {S{\o}gaard-Andersen}(2014)}]{Treuner-Lange_Sogaard-Andersen:2014}%
  \BibitemOpen
  \bibfield  {author} {\bibinfo {author} {\bibfnamefont {A.}~\bibnamefont
  {Treuner-Lange}}\ and\ \bibinfo {author} {\bibfnamefont {L.}~\bibnamefont
  {S{\o}gaard-Andersen}},\ }\href {\doibase 10.1083/jcb.201403136} {\bibfield
  {journal} {\bibinfo  {journal} {The Journal of Cell Biology}\ }\textbf
  {\bibinfo {volume} {206}},\ \bibinfo {pages} {7} (\bibinfo {year}
  {2014})}\BibitemShut {NoStop}%
\bibitem [{\citenamefont {Schuhmacher}\ \emph {et~al.}(2015)\citenamefont
  {Schuhmacher}, \citenamefont {Thormann}, \citenamefont {Bange},\ and\
  \citenamefont {Albers}}]{Schuhmacher_etal:2015}%
  \BibitemOpen
  \bibfield  {author} {\bibinfo {author} {\bibfnamefont {J.~S.}\ \bibnamefont
  {Schuhmacher}}, \bibinfo {author} {\bibfnamefont {K.~M.}\ \bibnamefont
  {Thormann}}, \bibinfo {author} {\bibfnamefont {G.}~\bibnamefont {Bange}}, \
  and\ \bibinfo {author} {\bibfnamefont {S.-V.}\ \bibnamefont {Albers}},\
  }\href {\doibase 10.1093/femsre/fuv034} {\ \textbf {\bibinfo {volume} {39}},\
  \bibinfo {pages} {812} (\bibinfo {year} {2015})}\BibitemShut {NoStop}%
\bibitem [{\citenamefont {Halatek}\ \emph {et~al.}(2018)\citenamefont
  {Halatek}, \citenamefont {Brauns},\ and\ \citenamefont
  {Frey}}]{halatek_brauns_frey:2018}%
  \BibitemOpen
  \bibfield  {author} {\bibinfo {author} {\bibfnamefont {J.}~\bibnamefont
  {Halatek}}, \bibinfo {author} {\bibfnamefont {F.}~\bibnamefont {Brauns}}, \
  and\ \bibinfo {author} {\bibfnamefont {E.}~\bibnamefont {Frey}},\ }\href
  {\doibase 10.1098/rstb.2017.0107} {\bibfield  {journal} {\bibinfo  {journal}
  {Philos Trans R Soc Lond B Biol Sci}\ } (\bibinfo {year} {2018}),\
  10.1098/rstb.2017.0107}\BibitemShut {NoStop}%
\bibitem [{Note1()}]{Note1}%
  \BibitemOpen
  \bibinfo {note} {Of course, such a process would also be limited by the
  duration of protein synthesis.}\BibitemShut {Stop}%
\bibitem [{\citenamefont {Segel}\ and\ \citenamefont
  {Jackson}(1972)}]{Segel_Jackson:1972}%
  \BibitemOpen
  \bibfield  {author} {\bibinfo {author} {\bibfnamefont {L.~A.}\ \bibnamefont
  {Segel}}\ and\ \bibinfo {author} {\bibfnamefont {J.~L.}\ \bibnamefont
  {Jackson}},\ }\href {\doibase 10.1016/0022-5193(72)90090-2} {\bibfield
  {journal} {\bibinfo  {journal} {Journal of Theoretical Biology}\ }\textbf
  {\bibinfo {volume} {37}},\ \bibinfo {pages} {545} (\bibinfo {year}
  {1972})}\BibitemShut {NoStop}%
\bibitem [{\citenamefont {Kondo}\ and\ \citenamefont
  {Miura}(2010)}]{Kondo_Miura:2010}%
  \BibitemOpen
  \bibfield  {author} {\bibinfo {author} {\bibfnamefont {S.}~\bibnamefont
  {Kondo}}\ and\ \bibinfo {author} {\bibfnamefont {T.}~\bibnamefont {Miura}},\
  }\href@noop {} {\bibfield  {journal} {\bibinfo  {journal} {Science (New York,
  NY)}\ }\textbf {\bibinfo {volume} {329}},\ \bibinfo {pages} {1616} (\bibinfo
  {year} {2010})}\BibitemShut {NoStop}%
\bibitem [{\citenamefont {Halatek}\ and\ \citenamefont
  {Frey}(2018)}]{halatek_frey:2018}%
  \BibitemOpen
  \bibfield  {author} {\bibinfo {author} {\bibfnamefont {J.}~\bibnamefont
  {Halatek}}\ and\ \bibinfo {author} {\bibfnamefont {E.}~\bibnamefont {Frey}},\
  }\href {\doibase 10.1038/s41567-017-0040-5} {\bibfield  {journal} {\bibinfo
  {journal} {Nature Physics}\ } (\bibinfo {year} {2018}),\
  10.1038/s41567-017-0040-5}\BibitemShut {NoStop}%
\bibitem [{\citenamefont {Huang}\ \emph {et~al.}(2003)\citenamefont {Huang},
  \citenamefont {Meir},\ and\ \citenamefont {Wingreen}}]{Huang_etal:2003}%
  \BibitemOpen
  \bibfield  {author} {\bibinfo {author} {\bibfnamefont {K.~C.}\ \bibnamefont
  {Huang}}, \bibinfo {author} {\bibfnamefont {Y.}~\bibnamefont {Meir}}, \ and\
  \bibinfo {author} {\bibfnamefont {N.~S.}\ \bibnamefont {Wingreen}},\ }\href
  {\doibase 10.1073/pnas.2135445100} {\bibfield  {journal} {\bibinfo  {journal}
  {Proceedings of the National Academy of Sciences of the United States of
  America}\ }\textbf {\bibinfo {volume} {100}},\ \bibinfo {pages} {12724}
  (\bibinfo {year} {2003})}\BibitemShut {NoStop}%
\bibitem [{\citenamefont {Halatek}\ and\ \citenamefont
  {Frey}(2012)}]{Halatek_Frey:2012}%
  \BibitemOpen
  \bibfield  {author} {\bibinfo {author} {\bibfnamefont {J.}~\bibnamefont
  {Halatek}}\ and\ \bibinfo {author} {\bibfnamefont {E.}~\bibnamefont {Frey}},\
  }\href {\doibase 10.1016/j.celrep.2012.04.005} {\bibfield  {journal}
  {\bibinfo  {journal} {Cell Reports}\ }\textbf {\bibinfo {volume} {1}},\
  \bibinfo {pages} {741} (\bibinfo {year} {2012})}\BibitemShut {NoStop}%
\bibitem [{\citenamefont {Meacci}\ \emph {et~al.}(2006)\citenamefont {Meacci},
  \citenamefont {Ries}, \citenamefont {Fischer-Friedrich}, \citenamefont
  {Kahya}, \citenamefont {Schwille},\ and\ \citenamefont
  {Kruse}}]{Meacci_etal:2006}%
  \BibitemOpen
  \bibfield  {author} {\bibinfo {author} {\bibfnamefont {G.}~\bibnamefont
  {Meacci}}, \bibinfo {author} {\bibfnamefont {J.}~\bibnamefont {Ries}},
  \bibinfo {author} {\bibfnamefont {E.}~\bibnamefont {Fischer-Friedrich}},
  \bibinfo {author} {\bibfnamefont {N.}~\bibnamefont {Kahya}}, \bibinfo
  {author} {\bibfnamefont {P.}~\bibnamefont {Schwille}}, \ and\ \bibinfo
  {author} {\bibfnamefont {K.}~\bibnamefont {Kruse}},\ }\href {\doibase
  10.1088/1478-3975/3/4/003} {\bibfield  {journal} {\bibinfo  {journal}
  {Physical biology}\ }\textbf {\bibinfo {volume} {3}},\ \bibinfo {pages} {255}
  (\bibinfo {year} {2006})}\BibitemShut {NoStop}%
\bibitem [{\citenamefont {Shih}\ \emph {et~al.}(2002)\citenamefont {Shih},
  \citenamefont {Fu}, \citenamefont {King}, \citenamefont {Le},\ and\
  \citenamefont {Rothfield}}]{Shih_etal:2002}%
  \BibitemOpen
  \bibfield  {author} {\bibinfo {author} {\bibfnamefont {Y.-L.}\ \bibnamefont
  {Shih}}, \bibinfo {author} {\bibfnamefont {X.}~\bibnamefont {Fu}}, \bibinfo
  {author} {\bibfnamefont {G.~F.}\ \bibnamefont {King}}, \bibinfo {author}
  {\bibfnamefont {T.}~\bibnamefont {Le}}, \ and\ \bibinfo {author}
  {\bibfnamefont {L.}~\bibnamefont {Rothfield}},\ }\href@noop {} {\bibfield
  {journal} {\bibinfo  {journal} {The EMBO journal}\ }\textbf {\bibinfo
  {volume} {21}},\ \bibinfo {pages} {3347} (\bibinfo {year}
  {2002})}\BibitemShut {NoStop}%
\bibitem [{Note2()}]{Note2}%
  \BibitemOpen
  \bibinfo {note} {In general, a given reaction-diffusion equation can generate
  a plethora of spatio-temporal patterns, as is well known from classical
  equations like the complex Ginzburg-Landau equation \cite
  {Aranson_Kramer:2002} or the Gray-Scott equation \cite {Gray_Scott:1983,
  Gray_Scott:1984, Gray_Scott:1985, Pearson:1993, Lee_etal:1993}. Conversely, a
  given pattern can be produced by a vast variety of mathematical equations.
  Hence, one must be careful to avoid falling into the trap: ``Cum hoc ergo
  propter hoc'' (correlation does not imply causation).}\BibitemShut {Stop}%
\bibitem [{Note3()}]{Note3}%
  \BibitemOpen
  \bibinfo {note} {It should be noted that the condition on the particle
  numbers mainly serves to emphasise the sequestration mechanism. In order for
  MinD to accumulate in polar zones the action of MinE must be disabled, and
  specifying that there are fewer MinE particles permits them to be spatially
  confined. Outside of this zone MinD can accumulate on the membrane. It has
  been speculated \cite {Halatek_Frey:2012} that other mechanisms, such as
  transient MinE membrane binding, might provide alternative ways to
  transiently disable the action of MinE, removing the requirement from the
  particle numbers. The exact mechanism needs to be investigated in future
  experiments as well as in the framework of theoretical models.}\BibitemShut
  {Stop}%
\bibitem [{\citenamefont {Wu}\ \emph {et~al.}(2015)\citenamefont {Wu},
  \citenamefont {van Schie}, \citenamefont {Keymer},\ and\ \citenamefont
  {Dekker}}]{Wu_etal:2015}%
  \BibitemOpen
  \bibfield  {author} {\bibinfo {author} {\bibfnamefont {F.}~\bibnamefont
  {Wu}}, \bibinfo {author} {\bibfnamefont {B.~G.~C.}\ \bibnamefont {van
  Schie}}, \bibinfo {author} {\bibfnamefont {J.~E.}\ \bibnamefont {Keymer}}, \
  and\ \bibinfo {author} {\bibfnamefont {C.}~\bibnamefont {Dekker}},\ }\href
  {\doibase 10.1038/nnano.2015.126} {\bibfield  {journal} {\bibinfo  {journal}
  {Nature Nanotechnology}\ }\textbf {\bibinfo {volume} {10}},\ \bibinfo {pages}
  {719} (\bibinfo {year} {2015})}\BibitemShut {NoStop}%
\bibitem [{\citenamefont {Shih}\ \emph {et~al.}(2005)\citenamefont {Shih},
  \citenamefont {Kawagishi},\ and\ \citenamefont {Rothfield}}]{Shih_etal:2005}%
  \BibitemOpen
  \bibfield  {author} {\bibinfo {author} {\bibfnamefont {Y.-L.}\ \bibnamefont
  {Shih}}, \bibinfo {author} {\bibfnamefont {I.}~\bibnamefont {Kawagishi}}, \
  and\ \bibinfo {author} {\bibfnamefont {L.}~\bibnamefont {Rothfield}},\ }\href
  {\doibase 10.1111/j.1365-2958.2005.04841.x} {\bibfield  {journal} {\bibinfo
  {journal} {Molecular Microbiology}\ }\textbf {\bibinfo {volume} {58}},\
  \bibinfo {pages} {917} (\bibinfo {year} {2005})}\BibitemShut {NoStop}%
\bibitem [{\citenamefont {Wu}\ \emph {et~al.}(2016)\citenamefont {Wu},
  \citenamefont {Halatek}, \citenamefont {Reiter}, \citenamefont {Kingma},
  \citenamefont {Frey},\ and\ \citenamefont {Dekker}}]{Wu_etal:2016}%
  \BibitemOpen
  \bibfield  {author} {\bibinfo {author} {\bibfnamefont {F.}~\bibnamefont
  {Wu}}, \bibinfo {author} {\bibfnamefont {J.}~\bibnamefont {Halatek}},
  \bibinfo {author} {\bibfnamefont {M.}~\bibnamefont {Reiter}}, \bibinfo
  {author} {\bibfnamefont {E.}~\bibnamefont {Kingma}}, \bibinfo {author}
  {\bibfnamefont {E.}~\bibnamefont {Frey}}, \ and\ \bibinfo {author}
  {\bibfnamefont {C.}~\bibnamefont {Dekker}},\ }\href {\doibase
  10.15252/MSB.20156724} {\bibfield  {journal} {\bibinfo  {journal} {Molecular
  Systems Biology}\ }\textbf {\bibinfo {volume} {12}},\ \bibinfo {pages} {642}
  (\bibinfo {year} {2016})}\BibitemShut {NoStop}%
\bibitem [{\citenamefont {Corbin}\ \emph {et~al.}(2002)\citenamefont {Corbin},
  \citenamefont {Yu},\ and\ \citenamefont {Margolin}}]{Corbin_etal:2002}%
  \BibitemOpen
  \bibfield  {author} {\bibinfo {author} {\bibfnamefont {B.~D.}\ \bibnamefont
  {Corbin}}, \bibinfo {author} {\bibfnamefont {X.-C.}\ \bibnamefont {Yu}}, \
  and\ \bibinfo {author} {\bibfnamefont {W.}~\bibnamefont {Margolin}},\ }\href
  {\doibase 10.1093/emboj/21.8.1998} {\bibfield  {journal} {\bibinfo  {journal}
  {The EMBO journal}\ }\textbf {\bibinfo {volume} {21}},\ \bibinfo {pages}
  {1998} (\bibinfo {year} {2002})}\BibitemShut {NoStop}%
\bibitem [{\citenamefont {Touhami}\ \emph {et~al.}(2006)\citenamefont
  {Touhami}, \citenamefont {Jericho},\ and\ \citenamefont
  {Rutenberg}}]{Touhami_etal:2006}%
  \BibitemOpen
  \bibfield  {author} {\bibinfo {author} {\bibfnamefont {A.}~\bibnamefont
  {Touhami}}, \bibinfo {author} {\bibfnamefont {M.}~\bibnamefont {Jericho}}, \
  and\ \bibinfo {author} {\bibfnamefont {A.~D.}\ \bibnamefont {Rutenberg}},\
  }\href {\doibase 10.1128/JB.00911-06} {\bibfield  {journal} {\bibinfo
  {journal} {Journal of Bacteriology}\ }\textbf {\bibinfo {volume} {188}},\
  \bibinfo {pages} {7661} (\bibinfo {year} {2006})}\BibitemShut {NoStop}%
\bibitem [{\citenamefont {Varma}\ \emph {et~al.}(2008)\citenamefont {Varma},
  \citenamefont {Huang},\ and\ \citenamefont {Young}}]{Varma_etal:2008}%
  \BibitemOpen
  \bibfield  {author} {\bibinfo {author} {\bibfnamefont {A.}~\bibnamefont
  {Varma}}, \bibinfo {author} {\bibfnamefont {K.~C.}\ \bibnamefont {Huang}}, \
  and\ \bibinfo {author} {\bibfnamefont {K.~D.}\ \bibnamefont {Young}},\ }\href
  {\doibase 10.1128/JB.00720-07} {\bibfield  {journal} {\bibinfo  {journal}
  {Journal of Bacteriology}\ }\textbf {\bibinfo {volume} {190}},\ \bibinfo
  {pages} {2106} (\bibinfo {year} {2008})}\BibitemShut {NoStop}%
\bibitem [{\citenamefont {M{\"a}nnik}\ \emph {et~al.}(2012)\citenamefont
  {M{\"a}nnik}, \citenamefont {Wu}, \citenamefont {Hol}, \citenamefont
  {Bisicchia}, \citenamefont {Sherratt}, \citenamefont {Keymer},\ and\
  \citenamefont {Dekker}}]{Maennik_etal:2012}%
  \BibitemOpen
  \bibfield  {author} {\bibinfo {author} {\bibfnamefont {J.}~\bibnamefont
  {M{\"a}nnik}}, \bibinfo {author} {\bibfnamefont {F.}~\bibnamefont {Wu}},
  \bibinfo {author} {\bibfnamefont {F.~J.~H.}\ \bibnamefont {Hol}}, \bibinfo
  {author} {\bibfnamefont {P.}~\bibnamefont {Bisicchia}}, \bibinfo {author}
  {\bibfnamefont {D.~J.}\ \bibnamefont {Sherratt}}, \bibinfo {author}
  {\bibfnamefont {J.~E.}\ \bibnamefont {Keymer}}, \ and\ \bibinfo {author}
  {\bibfnamefont {C.}~\bibnamefont {Dekker}},\ }\href {\doibase
  10.1073/pnas.1120854109} {\bibfield  {journal} {\bibinfo  {journal}
  {Proceedings of the National Academy of Sciences of the United States of
  America}\ }\textbf {\bibinfo {volume} {109}},\ \bibinfo {pages} {6957}
  (\bibinfo {year} {2012})}\BibitemShut {NoStop}%
\bibitem [{\citenamefont {Fange}\ and\ \citenamefont
  {Elf}(2006)}]{Fange_Elf:2006}%
  \BibitemOpen
  \bibfield  {author} {\bibinfo {author} {\bibfnamefont {D.}~\bibnamefont
  {Fange}}\ and\ \bibinfo {author} {\bibfnamefont {J.}~\bibnamefont {Elf}},\
  }\href {\doibase 10.1371/journal.pcbi.0020080} {\bibfield  {journal}
  {\bibinfo  {journal} {Plos Computational Biology}\ }\textbf {\bibinfo
  {volume} {2}},\ \bibinfo {pages} {e80} (\bibinfo {year} {2006})}\BibitemShut
  {NoStop}%
\bibitem [{\citenamefont {Renner}\ and\ \citenamefont
  {Weibel}(2012)}]{Renner_Weibel:2012}%
  \BibitemOpen
  \bibfield  {author} {\bibinfo {author} {\bibfnamefont {L.~D.}\ \bibnamefont
  {Renner}}\ and\ \bibinfo {author} {\bibfnamefont {D.~B.}\ \bibnamefont
  {Weibel}},\ }\href {\doibase 10.1074/jbc.M112.407817} {\bibfield  {journal}
  {\bibinfo  {journal} {The Journal of biological chemistry}\ }\textbf
  {\bibinfo {volume} {287}},\ \bibinfo {pages} {38835} (\bibinfo {year}
  {2012})}\BibitemShut {NoStop}%
\bibitem [{Note4()}]{Note4}%
  \BibitemOpen
  \bibinfo {note} {This is surprising, because Turing instabilities are
  generically associated with the existence of a \protect \textit
  {characteristic} (or intrinsic) wave length in the literature. This is
  evidently not the case here.}\BibitemShut {Stop}%
\bibitem [{\citenamefont {Otsuji}\ \emph {et~al.}(2007)\citenamefont {Otsuji},
  \citenamefont {Ishihara}, \citenamefont {Co}, \citenamefont {Kaibuchi},
  \citenamefont {Mochizuki},\ and\ \citenamefont {Kuroda}}]{Otsuji_etal:2007}%
  \BibitemOpen
  \bibfield  {author} {\bibinfo {author} {\bibfnamefont {M.}~\bibnamefont
  {Otsuji}}, \bibinfo {author} {\bibfnamefont {S.}~\bibnamefont {Ishihara}},
  \bibinfo {author} {\bibfnamefont {C.}~\bibnamefont {Co}}, \bibinfo {author}
  {\bibfnamefont {K.}~\bibnamefont {Kaibuchi}}, \bibinfo {author}
  {\bibfnamefont {A.}~\bibnamefont {Mochizuki}}, \ and\ \bibinfo {author}
  {\bibfnamefont {S.}~\bibnamefont {Kuroda}},\ }\href@noop {} {\bibfield
  {journal} {\bibinfo  {journal} {PLoS Comput. Biol.}\ }\textbf {\bibinfo
  {volume} {3}},\ \bibinfo {pages} {e108} (\bibinfo {year} {2007})}\BibitemShut
  {NoStop}%
\bibitem [{Note5()}]{Note5}%
  \BibitemOpen
  \bibinfo {note} {In 1966 Mark Kac published an article entitled ``Can one
  hear the shape of a drum?''\cite {Kac:1966}. As the dynamics (frequency
  spectrum) of an elastic membrane whose boundary is clamped is described by
  the Helmholtz equation $\nabla ^2 u + \sigma u =0$ with Dirichlet boundaries,
  $\nabla u \mid _\perp = 0$, this amounts to asking how strongly the
  eigenvalues $\sigma $ depend on the shape of the domain boundary. Here we ask
  a much more intricate question, as the dynamics of pattern forming systems
  are nonlinear and we would like to know the nonlinear attractor for a given
  shape and size of a cell.}\BibitemShut {Stop}%
\bibitem [{\citenamefont {Thalmeier}\ \emph {et~al.}(2016)\citenamefont
  {Thalmeier}, \citenamefont {Halatek},\ and\ \citenamefont
  {Frey}}]{Thalmeier_etal:2016}%
  \BibitemOpen
  \bibfield  {author} {\bibinfo {author} {\bibfnamefont {D.}~\bibnamefont
  {Thalmeier}}, \bibinfo {author} {\bibfnamefont {J.}~\bibnamefont {Halatek}},
  \ and\ \bibinfo {author} {\bibfnamefont {E.}~\bibnamefont {Frey}},\ }\href
  {\doibase 10.1073/pnas.1515191113} {\bibfield  {journal} {\bibinfo  {journal}
  {Proceedings of the National Academy of Sciences of the United States of
  America}\ }\textbf {\bibinfo {volume} {113}},\ \bibinfo {pages} {548}
  (\bibinfo {year} {2016})}\BibitemShut {NoStop}%
\bibitem [{\citenamefont {Loose}\ \emph {et~al.}(2008)\citenamefont {Loose},
  \citenamefont {Fischer-Friedrich}, \citenamefont {Ries}, \citenamefont
  {Kruse},\ and\ \citenamefont {Schwille}}]{Loose_etal:2008}%
  \BibitemOpen
  \bibfield  {author} {\bibinfo {author} {\bibfnamefont {M.}~\bibnamefont
  {Loose}}, \bibinfo {author} {\bibfnamefont {E.}~\bibnamefont
  {Fischer-Friedrich}}, \bibinfo {author} {\bibfnamefont {J.}~\bibnamefont
  {Ries}}, \bibinfo {author} {\bibfnamefont {K.}~\bibnamefont {Kruse}}, \ and\
  \bibinfo {author} {\bibfnamefont {P.}~\bibnamefont {Schwille}},\ }\href
  {\doibase 10.1126/science.1154413} {\bibfield  {journal} {\bibinfo  {journal}
  {Science (New York, NY)}\ }\textbf {\bibinfo {volume} {320}},\ \bibinfo
  {pages} {789} (\bibinfo {year} {2008})}\BibitemShut {NoStop}%
\bibitem [{\citenamefont {Caspi}\ and\ \citenamefont
  {Dekker}(2016)}]{Caspi_Dekker:2016}%
  \BibitemOpen
  \bibfield  {author} {\bibinfo {author} {\bibfnamefont {Y.}~\bibnamefont
  {Caspi}}\ and\ \bibinfo {author} {\bibfnamefont {C.}~\bibnamefont {Dekker}},\
  }\href {\doibase 10.7554/eLife.19271} {\bibfield  {journal} {\bibinfo
  {journal} {eLife}\ }\textbf {\bibinfo {volume} {5}},\ \bibinfo {pages}
  {e19271} (\bibinfo {year} {2016})}\BibitemShut {NoStop}%
\bibitem [{\citenamefont {Loose}\ \emph
  {et~al.}(2011{\natexlab{b}})\citenamefont {Loose}, \citenamefont
  {Fischer-Friedrich}, \citenamefont {Herold}, \citenamefont {Kruse},\ and\
  \citenamefont {Schwille}}]{Loose_etal:2011}%
  \BibitemOpen
  \bibfield  {author} {\bibinfo {author} {\bibfnamefont {M.}~\bibnamefont
  {Loose}}, \bibinfo {author} {\bibfnamefont {E.}~\bibnamefont
  {Fischer-Friedrich}}, \bibinfo {author} {\bibfnamefont {C.}~\bibnamefont
  {Herold}}, \bibinfo {author} {\bibfnamefont {K.}~\bibnamefont {Kruse}}, \
  and\ \bibinfo {author} {\bibfnamefont {P.}~\bibnamefont {Schwille}},\ }\href
  {\doibase 10.1038/nsmb.2037} {\bibfield  {journal} {\bibinfo  {journal}
  {Nature structural {\&} molecular biology}\ }\textbf {\bibinfo {volume}
  {18}},\ \bibinfo {pages} {577} (\bibinfo {year}
  {2011}{\natexlab{b}})}\BibitemShut {NoStop}%
\bibitem [{\citenamefont {Schweizer}\ \emph {et~al.}(2012)\citenamefont
  {Schweizer}, \citenamefont {Loose}, \citenamefont {Bonny}, \citenamefont
  {Kruse}, \citenamefont {M{\"o}nch},\ and\ \citenamefont
  {Schwille}}]{Schweizer_etal:2012}%
  \BibitemOpen
  \bibfield  {author} {\bibinfo {author} {\bibfnamefont {J.}~\bibnamefont
  {Schweizer}}, \bibinfo {author} {\bibfnamefont {M.}~\bibnamefont {Loose}},
  \bibinfo {author} {\bibfnamefont {M.}~\bibnamefont {Bonny}}, \bibinfo
  {author} {\bibfnamefont {K.}~\bibnamefont {Kruse}}, \bibinfo {author}
  {\bibfnamefont {I.}~\bibnamefont {M{\"o}nch}}, \ and\ \bibinfo {author}
  {\bibfnamefont {P.}~\bibnamefont {Schwille}},\ }\href {\doibase
  10.1073/pnas.1206953109} {\bibfield  {journal} {\bibinfo  {journal}
  {Proceedings of the National Academy of Sciences}\ }\textbf {\bibinfo
  {volume} {109}},\ \bibinfo {pages} {15283} (\bibinfo {year}
  {2012})}\BibitemShut {NoStop}%
\bibitem [{\citenamefont {Zieske}\ and\ \citenamefont
  {Schwille}(2013)}]{Zieske_Schwille:2013}%
  \BibitemOpen
  \bibfield  {author} {\bibinfo {author} {\bibfnamefont {K.}~\bibnamefont
  {Zieske}}\ and\ \bibinfo {author} {\bibfnamefont {P.}~\bibnamefont
  {Schwille}},\ }\href {\doibase 10.1002/anie.201207078} {\bibfield  {journal}
  {\bibinfo  {journal} {Angewandte Chemie (International ed. in English)}\
  }\textbf {\bibinfo {volume} {52}},\ \bibinfo {pages} {459} (\bibinfo {year}
  {2013})}\BibitemShut {NoStop}%
\bibitem [{\citenamefont {Zieske}\ \emph {et~al.}(2016)\citenamefont {Zieske},
  \citenamefont {Chwastek},\ and\ \citenamefont {Schwille}}]{Zieske_etal:2016}%
  \BibitemOpen
  \bibfield  {author} {\bibinfo {author} {\bibfnamefont {K.}~\bibnamefont
  {Zieske}}, \bibinfo {author} {\bibfnamefont {G.}~\bibnamefont {Chwastek}}, \
  and\ \bibinfo {author} {\bibfnamefont {P.}~\bibnamefont {Schwille}},\ }\href
  {\doibase 10.1002/ange.201606069} {\bibfield  {journal} {\bibinfo  {journal}
  {Angewandte Chemie}\ }\textbf {\bibinfo {volume} {128}},\ \bibinfo {pages}
  {13653} (\bibinfo {year} {2016})}\BibitemShut {NoStop}%
\bibitem [{\citenamefont {Ivanov}\ and\ \citenamefont
  {Mizuuchi}(2010)}]{Ivanov_Mizuuchi:2010}%
  \BibitemOpen
  \bibfield  {author} {\bibinfo {author} {\bibfnamefont {V.}~\bibnamefont
  {Ivanov}}\ and\ \bibinfo {author} {\bibfnamefont {K.}~\bibnamefont
  {Mizuuchi}},\ }\href {\doibase 10.1073/pnas.0911036107} {\bibfield  {journal}
  {\bibinfo  {journal} {Proceedings of the National Academy of Sciences of the
  United States of America}\ }\textbf {\bibinfo {volume} {107}},\ \bibinfo
  {pages} {8071} (\bibinfo {year} {2010})}\BibitemShut {NoStop}%
\bibitem [{\citenamefont {Zieske}\ and\ \citenamefont
  {Schwille}(2014)}]{Zieske_Schwille:2014}%
  \BibitemOpen
  \bibfield  {author} {\bibinfo {author} {\bibfnamefont {K.}~\bibnamefont
  {Zieske}}\ and\ \bibinfo {author} {\bibfnamefont {P.}~\bibnamefont
  {Schwille}},\ }\href {\doibase 10.7554/eLife.03949} {\bibfield  {journal}
  {\bibinfo  {journal} {eLife}\ }\textbf {\bibinfo {volume} {3}} (\bibinfo
  {year} {2014}),\ 10.7554/eLife.03949}\BibitemShut {NoStop}%
\bibitem [{Note6()}]{Note6}%
  \BibitemOpen
  \bibinfo {note} {We note that travelling wave patterns have also been
  observed \protect \textit {in vivo} \cite {Bonny_etal:2013}, albeit only upon
  massive over-expression of MinD and MinE, leading to highly elevated
  intracellular protein densities and pathological phenomenology \cite
  {Sliusarenko_etal:2011} relative to the wild type. While the exact protein
  densities in the experiments have not been measured, this observation is
  consistent with the observation of travelling waves in fully confined
  compartments, where the protein densities inside microfluidic chambers were
  also elevated \cite {Caspi_Dekker:2016}. For further discussion of the effect
  of protein densities we refer the reader to section \protect \ref
  {sec:polychotomy}.}\BibitemShut {Stop}%
\bibitem [{\citenamefont {Vecchiarelli}\ \emph {et~al.}(2016)\citenamefont
  {Vecchiarelli}, \citenamefont {Li}, \citenamefont {Mizuuchi}, \citenamefont
  {Hwang}, \citenamefont {Seol}, \citenamefont {Neuman},\ and\ \citenamefont
  {Mizuuchi}}]{Vecchiarelli_etal:2016}%
  \BibitemOpen
  \bibfield  {author} {\bibinfo {author} {\bibfnamefont {A.~G.}\ \bibnamefont
  {Vecchiarelli}}, \bibinfo {author} {\bibfnamefont {M.}~\bibnamefont {Li}},
  \bibinfo {author} {\bibfnamefont {M.}~\bibnamefont {Mizuuchi}}, \bibinfo
  {author} {\bibfnamefont {L.~C.}\ \bibnamefont {Hwang}}, \bibinfo {author}
  {\bibfnamefont {Y.}~\bibnamefont {Seol}}, \bibinfo {author} {\bibfnamefont
  {K.~C.}\ \bibnamefont {Neuman}}, \ and\ \bibinfo {author} {\bibfnamefont
  {K.}~\bibnamefont {Mizuuchi}},\ }\href {\doibase 10.1073/pnas.1600644113}
  {\bibfield  {journal} {\bibinfo  {journal} {Proceedings of the National
  Academy of Sciences of the United States of America}\ }\textbf {\bibinfo
  {volume} {113}},\ \bibinfo {pages} {E1479} (\bibinfo {year}
  {2016})}\BibitemShut {NoStop}%
\bibitem [{\citenamefont {Gray}\ and\ \citenamefont
  {Scott}(1983)}]{Gray_Scott:1983}%
  \BibitemOpen
  \bibfield  {author} {\bibinfo {author} {\bibfnamefont {P.}~\bibnamefont
  {Gray}}\ and\ \bibinfo {author} {\bibfnamefont {S.~K.}\ \bibnamefont
  {Scott}},\ }\href {\doibase 10.1016/0009-2509(83)80132-8} {\bibfield
  {journal} {\bibinfo  {journal} {Chemical Engineering Science}\ }\textbf
  {\bibinfo {volume} {38}},\ \bibinfo {pages} {29} (\bibinfo {year}
  {1983})}\BibitemShut {NoStop}%
\bibitem [{\citenamefont {Gray}\ and\ \citenamefont
  {Scott}(1984)}]{Gray_Scott:1984}%
  \BibitemOpen
  \bibfield  {author} {\bibinfo {author} {\bibfnamefont {P.}~\bibnamefont
  {Gray}}\ and\ \bibinfo {author} {\bibfnamefont {S.~K.}\ \bibnamefont
  {Scott}},\ }\href {\doibase 10.1016/0009-2509(84)87017-7} {\bibfield
  {journal} {\bibinfo  {journal} {Chemical Engineering Science}\ }\textbf
  {\bibinfo {volume} {39}},\ \bibinfo {pages} {1087} (\bibinfo {year}
  {1984})}\BibitemShut {NoStop}%
\bibitem [{\citenamefont {Gray}\ and\ \citenamefont
  {Scott}(1985)}]{Gray_Scott:1985}%
  \BibitemOpen
  \bibfield  {author} {\bibinfo {author} {\bibfnamefont {P.}~\bibnamefont
  {Gray}}\ and\ \bibinfo {author} {\bibfnamefont {S.~K.}\ \bibnamefont
  {Scott}},\ }\href {\doibase 10.1021/j100247a009} {\bibfield  {journal}
  {\bibinfo  {journal} {The Journal of Physical Chemistry}\ }\textbf {\bibinfo
  {volume} {89}},\ \bibinfo {pages} {22} (\bibinfo {year} {1985})}\BibitemShut
  {NoStop}%
\bibitem [{\citenamefont {Pearson}(1993)}]{Pearson:1993}%
  \BibitemOpen
  \bibfield  {author} {\bibinfo {author} {\bibfnamefont {J.~E.}\ \bibnamefont
  {Pearson}},\ }\href {\doibase 10.1126/science.261.5118.189} {\bibfield
  {journal} {\bibinfo  {journal} {Science (New York, NY)}\ }\textbf {\bibinfo
  {volume} {261}},\ \bibinfo {pages} {189} (\bibinfo {year}
  {1993})}\BibitemShut {NoStop}%
\bibitem [{\citenamefont {Lee}\ \emph {et~al.}(1993)\citenamefont {Lee},
  \citenamefont {McCormick}, \citenamefont {Ouyang},\ and\ \citenamefont
  {Swinney}}]{Lee_etal:1993}%
  \BibitemOpen
  \bibfield  {author} {\bibinfo {author} {\bibfnamefont {K.~J.}\ \bibnamefont
  {Lee}}, \bibinfo {author} {\bibfnamefont {W.~D.}\ \bibnamefont {McCormick}},
  \bibinfo {author} {\bibfnamefont {Q.}~\bibnamefont {Ouyang}}, \ and\ \bibinfo
  {author} {\bibfnamefont {H.~L.}\ \bibnamefont {Swinney}},\ }\href {\doibase
  10.1126/science.261.5118.192} {\bibfield  {journal} {\bibinfo  {journal}
  {Science (New York, NY)}\ }\textbf {\bibinfo {volume} {261}},\ \bibinfo
  {pages} {192} (\bibinfo {year} {1993})}\BibitemShut {NoStop}%
\bibitem [{Note7()}]{Note7}%
  \BibitemOpen
  \bibinfo {note} {Either directly, or by complex formation as for MinDE
  complexes.}\BibitemShut {Stop}%
\bibitem [{Note8()}]{Note8}%
  \BibitemOpen
  \bibinfo {note} {Assuming a cylindrical geometry for simplicity, the volume
  to surface ratio is $\sim r/2$, i.e.\/ well below $1 \protect \tmspace
  +\thinmuskip {.1667em}\mu $m for typical cell radii $r$.}\BibitemShut {Stop}%
\bibitem [{\citenamefont {Halatek}\ and\ \citenamefont
  {Frey}(2014)}]{Halatek_Frey:2014}%
  \BibitemOpen
  \bibfield  {author} {\bibinfo {author} {\bibfnamefont {J.}~\bibnamefont
  {Halatek}}\ and\ \bibinfo {author} {\bibfnamefont {E.}~\bibnamefont {Frey}},\
  }\href {\doibase 10.1073/pnas.1220971111} {\bibfield  {journal} {\bibinfo
  {journal} {Proceedings of the National Academy of Sciences}\ }\textbf
  {\bibinfo {volume} {111}},\ \bibinfo {pages} {E1817} (\bibinfo {year}
  {2014})}\BibitemShut {NoStop}%
\bibitem [{\citenamefont {Aranson}\ and\ \citenamefont
  {Kramer}(2002)}]{Aranson_Kramer:2002}%
  \BibitemOpen
  \bibfield  {author} {\bibinfo {author} {\bibfnamefont {I.}~\bibnamefont
  {Aranson}}\ and\ \bibinfo {author} {\bibfnamefont {L.}~\bibnamefont
  {Kramer}},\ }\href {http://rmp.aps.org/abstract/RMP/v74/i1/p99{\_}1}
  {\bibfield  {journal} {\bibinfo  {journal} {Reviews of Modern Physics}\
  }\textbf {\bibinfo {volume} {74}},\ \bibinfo {pages} {99} (\bibinfo {year}
  {2002})}\BibitemShut {NoStop}%
\bibitem [{\citenamefont {Denk}\ \emph {et~al.}(2018)\citenamefont {Denk},
  \citenamefont {Kretschmer}, \citenamefont {Halatek}, \citenamefont {Hartl},
  \citenamefont {Schwille},\ and\ \citenamefont {Frey}}]{denk_etal:2018}%
  \BibitemOpen
  \bibfield  {author} {\bibinfo {author} {\bibfnamefont {J.}~\bibnamefont
  {Denk}}, \bibinfo {author} {\bibfnamefont {S.}~\bibnamefont {Kretschmer}},
  \bibinfo {author} {\bibfnamefont {J.}~\bibnamefont {Halatek}}, \bibinfo
  {author} {\bibfnamefont {C.}~\bibnamefont {Hartl}}, \bibinfo {author}
  {\bibfnamefont {P.}~\bibnamefont {Schwille}}, \ and\ \bibinfo {author}
  {\bibfnamefont {E.}~\bibnamefont {Frey}},\ }\href@noop {} {\bibfield
  {journal} {\bibinfo  {journal} {(to be published)}\ } (\bibinfo {year}
  {2018})}\BibitemShut {NoStop}%
\end{thebibliography}
%

\end{document}